\documentclass[twocolumn,aps,showpacs,superscriptaddress,pre]{revtex4}

\newcommand{\ca}{x}  
\newcommand{\ff}{y}  
\newcommand{\AF}{{\cal A}F} 
\newcommand{\re}{\rho_f}  
\newcommand{\rt}{\rho_t}  
\newcommand{\rs}{\rho_s}  
\newcommand{\fa}{{\mathfrak n}}  
\newcommand{\Sf}{S_f}  
\newcommand{\Y}{{\bf J8}}
\newcommand{\Z}{{\bf J4}}

\newcommand{\g}{j}

\newcommand{\wn}{\tau}
\newcommand{\F}{{\tilde F}}

\newcommand{\wt}{Q}

\newcommand{\bb}{\beta}

\newcommand{\W}{{\cal W}}   

\newcommand{\cJ}{\cal J}

\usepackage{amsfonts}
\usepackage{amsmath,amssymb,graphicx}
\usepackage{algorithmic}
\usepackage{color}

\begin{document}
\title{Population annealing: Theory and application in spin glasses} 

\author{Wenlong Wang}
\email{wenlong@physics.umass.edu}
\affiliation{Department of Physics, University of Massachusetts,
Amherst, Massachusetts 01003 USA}

\author{Jonathan Machta}
\email{machta@physics.umass.edu}
\affiliation{Department of Physics, University of Massachusetts,
Amherst, Massachusetts 01003 USA}
\affiliation{Santa Fe Institute, 1399 Hyde Park Road, Santa Fe, New 
Mexico 87501 USA}

\author{Helmut G. Katzgraber}
\affiliation{Department of Physics and Astronomy, Texas A\&M University,
College Station, Texas 77843-4242, USA}
\affiliation{Materials Science and Engineering, Texas A\&M University,
College Station, Texas 77843, USA}
\affiliation{Santa Fe Institute, 1399 Hyde Park Road, Santa Fe, New
Mexico 87501 USA}
\affiliation{Applied Mathematics Research Centre, Coventry University,
Coventry, CV1 5FB, England}

\begin{abstract}

Population annealing is an efficient sequential Monte Carlo algorithm
for simulating equilibrium states of systems with rough free energy
landscapes.  The theory of population annealing is presented, and
systematic and statistical errors are discussed.  The behavior of the
algorithm is studied in the context of large-scale simulations of the
three-dimensional Ising spin glass and the performance of the algorithm
is compared to parallel tempering.  It is found that the two algorithms
are similar in efficiency though with different strengths and
weaknesses.

\end{abstract}

\maketitle

\section{Introduction}

One of the most difficult challenges faced in computational physics is
simulating the equilibrium states of disordered systems with rough free
energy landscapes, such as spin glasses \cite{binder:86,stein:13}.
Standard Markov-chain Monte Carlo methods operating at a single
temperature, such as the Metropolis algorithm, equilibrate extremely
slowly at low temperatures due to trapping in metastable states.
Algorithms that make use of simulations at many temperatures partially
solve this problem and are now the methods of choice for equilibrium
simulations of disordered systems with frustration.  Markov-chain Monte
Carlo methods in this general category include parallel tempering Monte
Carlo \cite{swendsen:86,geyer:91,hukushima:96,earl:05,katzgraber:06a},
simulated tempering Monte Carlo \cite{marinari:92}, and the Wang-Landau
algorithm \cite{wang:01,wang:01a}. Population annealing Monte Carlo
\cite{hukushima:03,machta:10,machta:11}, the topic of this paper, is an
alternative to these multicanonical algorithms.  Population annealing
also employs many temperatures. However, it is not a Markov-chain Monte
Carlo method.  Instead, population annealing is a {\em sequential} Monte
Carlo method \cite{doucet:01}.  Population annealing was introduced by
Hukushima and Iba \cite{hukushima:03} and further developed in
Refs.~\cite{machta:10} and \cite{machta:11}. The algorithm was
independently discovered and called ``sequential Monte Carlo simulated
annealing" in Ref.~\cite{zhou:10}. Sequential Monte Carlo algorithms are
not commonly used in computational physics. However, the approach has
been applied to some problems, e.g., the diffusion Monte Carlo method
for finding ground states of many-body Schrodinger equations and
Grassberger's ``Go with Winners'' method \cite{grass:02}.

Population annealing has been successfully used in large-scale
simulations of Ising spin glasses by the present authors
\cite{wang:14,wang:15a}. One of the purposes of this paper is to give
additional details of the implementation and performance of the
population annealing algorithm as used in Refs.~\cite{wang:14} and
\cite{wang:15a}. We have also successfully used population annealing as
a heuristic to find spin-glass ground states \cite{wang:15} and have
shown that population annealing is comparably efficient to parallel
tempering Monte Carlo while both are far more efficient than simulated
annealing \cite{kirkpatrick:83}.  The current state-of-the-art algorithm
for simulations of spin glasses and related systems with rough
free-energy landscapes is parallel tempering Monte Carlo. Another
purpose of this paper to argue that population annealing is a useful
alternative and potentially superior to parallel tempering for
large-scale studies of the equilibrium properties of spin glasses and
other disordered systems.  We carry out a detailed comparison of
population annealing and parallel tempering for simulating spin glasses
and we find that the two methods are comparably efficient for sampling
thermal states, although each method has advantages and disadvantages.
We also develop a theory that quantifies the rate of convergence to
equilibrium of population annealing and compare the theoretical
predictions to simulations.

The outline of the paper is as follows. In Sec.~\ref{sec:papt} we
introduce both population annealing and parallel tempering Monte Carlo,
and in Sec.~\ref{sec:pafeatures} we discuss several features of
population annealing.  Section \ref{sec:error} is concerned with the
systematic and statistical errors in population annealing and the
section concludes with a comparison of errors to those in Markov chain
Monte Carlo methods such as parallel tempering.  Section
\ref{sec:modelmethods} introduces the Edwards-Anderson Ising spin glass
model and gives details of the simulations  and the quantities that were
measured. Section \ref{sec:results} presents the results of large-scale
simulations using both population annealing and parallel tempering with
an emphasis on elucidating the properties of population annealing and
comparing them to parallel tempering.  The paper concludes with a
discussion in Sec.~\ref{sec:discussion}

\section{Population Annealing and Parallel Tempering Monte Carlo}
\label{sec:papt}

\subsection{Population annealing}

Population annealing (PA) is closely related to simulated
annealing~\cite{kirkpatrick:83} (SA) except that it uses a population of
replicas and this population is resampled at each temperature step.
Like simulated annealing, PA involves lowering the temperature of the
system through a sequence of temperatures from a high temperature where
equilibration (also known as thermalization) is easy to a low, target
temperature $T_0$ where equilibration is difficult. Unlike SA, PA is
designed to simulate the equilibrium Gibbs distribution at each
temperature that is traversed. The resampling step ensures that the
population stays close to the equilibrium ensemble.  Just as in
simulated annealing, at each temperature, each replica is acted on by a
Markov-chain Monte Carlo (MCMC) procedure such as the Metropolis
algorithm. The annealing schedule consists of a sequence
$\{\bb_{N_T-1},\ldots, \bb_{0} \} $ of $N_T$ inverse temperatures,
$\bb=1/T$ labeled in descending order in $\bb$ so that
$\bb_{j+1}<\bb_j$. In our studies, the inverse temperatures are equally
spaced, starting at infinite temperature, $\bb_{N_T-1}=0$ and ending at
$\bb_0=5$. The MCMC method is the Metropolis algorithm
\cite{metropolis:49,metropolis:53}, which is applied for $N_S$ sweeps at
each temperature. The initial population size is $R$ and each replica is
independently initialized with random spins corresponding to an infinite
temperature ensemble.  In our implementation the population size
fluctuates. In a given run at inverse temperature $\beta$ the population
size is $\tilde{R}_\beta$ with a mean value of $R$.

The resampling step uses differential reproduction of replicas with a
number of copies depending on the replica's energy. Let $E_\g$ be the
energy of replica $\g$. Given an equilibrium population at $\beta$, the
goal is to resample the population so that it is an equilibrium
population at $\beta^\prime$ with $\beta^\prime > \beta$.  For a replica
$\g$, with energy $E_\g$ the ratio of the statistical weights at $\beta$
and $\beta^\prime$ is $\exp\left[ -(\beta^\prime-\beta)E_\g\right]$.
Note that this is the reweighting used in the histogram re-weighting
method \cite{ferrenberg:88}.

Since the typical energy is large and negative, the reweighting factor
is much greater than unity so that a normalization is needed to keep the
population size close to $R$. The normalized  weights
$\wn_\g(\beta,\beta^\prime)$ are given by,
\begin{equation}
\label{eq:wn}
\wn_\g(\beta,\beta^\prime)
=\frac{R}{\tilde{R}_\beta} \cdot
\frac{e^{-(\beta^\prime-\beta)E_{\g}}}{\wt(\beta,\beta^\prime)} ,
\end{equation}
where $\wt(\beta,\beta^\prime)$ is the normalization,
\begin{equation}
\label{eq:Q}
\wt(\beta,\beta^\prime)
=\frac{1}{\tilde{R}_\beta}
\sum_{j=1}^{\tilde{R}_\beta} 
e^{-(\beta^\prime-\beta)E_{\g}} .
\end{equation}
Note that the sum of $\wn_\g(\beta,\beta^\prime)$ over $\g$ is $R$.

The new population at temperature $\beta^\prime$ is obtained by
resampling the old population such that the number of copies $n_{\g}$ of
replica $\g$ is a random non-negative integer whose mean is
$\wn_\g(\beta,\beta^\prime)$. There are many ways to choose $n_{\g}$ to
satisfy this requirement~\cite{douc:05}.  Some of these methods such as
a multinomially distributed $n_{\g}$ keep the population size fixed
while others allow the population size to fluctuate.  We choose a method
that minimizes the variance of $n_{\g}$ and has $\sqrt{R}$ fluctuations
in the population size.  Let $n_{\g}$ be $\lfloor \wn_\g \rfloor$ with
probability $1-(\wn_\g-\lfloor \wn_\g \rfloor)$ or $\lceil \wn_\g
\rceil$  with probability $(\wn_\g-\lfloor \wn_\g \rfloor)$ where
$\lfloor \wn_\g \rfloor$ is the floor (greatest integer less than)
$\wn_\g$ and $\lceil \wn_\g \rceil$ is the ceiling (least integer
greater than) $\wn_\g$.  By minimizing the variance of  $n_{\g}$ we
reduce correlations in the population while the fluctuating population
size creates a small overhead in memory usage.

Consider a single temperature step in PA. If the original, higher
temperature population is an equilibrium ensemble representing the Gibbs
distribution at inverse temperature $\beta$ then the final, lower
temperature population is also an equilibrium ensemble at inverse
temperature $\beta^\prime$.  However, the new population is correlated
due to copying replicas in the resampling step and, for finite $R$, the
new population represents a biased ensemble due to a lack of
representation of the low energy tail of the $\beta^\prime$ Gibbs
distribution.  These errors are partially corrected by the MCMC sweeps
$\beta^\prime$.  Statistical and systematic errors are discussed in
Sec.~\ref{sec:error}.

The full PA algorithm is a sequence of $N_T-1$ annealing steps starting
from infinite temperature. In each annealing step the population is
resampled and then $N_S$ sweeps of the Metropolis algorithm are carried
out on each replica.

\subsection{Parallel Tempering}

In this section, we briefly describe parallel tempering (PT) Monte Carlo,
the state-of-the-art Monte Carlo algorithm for spin glasses and many
other frustrated systems. Parallel tempering is a Markov-chain Monte
Carlo algorithm while population annealing is a sequential Monte Carlo
algorithm, nonetheless, the two algorithms share many similarities.  In
parallel tempering, a set of $N_T$ temperatures is used ranging from a
high temperature that is easy to equilibrate to a low temperature of
interest, $T_0$. There is a single replica of the system  at each of
these temperatures and each of these replicas is operated on by a MCMC
method such as the Metropolis algorithm at that temperature. After $N_S$
sweeps of the replicas at their respective temperatures, replica
exchange moves are proposed.  In a replica exchange move, replicas at
two temperatures are proposed for swapping.  Typically, the two
temperatures are chosen to be neighboring temperatures in the list of
temperatures. Let these two inverse temperatures be $\bb$ and
$\bb^\prime$ with $\bb^\prime > \bb$ and let $E$ and $E^\prime$ be the
respective energies of the replicas at these two temperatures.  The swap
is accepted with probability $\min\left[1, \exp[(\bb^\prime - \bb) (
E^\prime - E) ]\right]$. It is easily shown that this swap probability
satisfies detailed balance \cite{hukushima:96} with respect to the
product measure of Gibbs distributions at the $N_T$ temperatures so that
PT converges to equilibrium at each temperature. Diffusion of replicas
from low temperature to high temperature and back, called `round trips'
allows PT to surmount barriers in the free energy
landscape~\cite{katzgraber:06a,machta:09}.

\section{Features of population annealing}
\label{sec:pafeatures}

\subsection{Free energy estimator}

The free-energy difference between the highest and lowest temperature is
easily measured using population annealing \cite{machta:10}. If the
highest temperature is infinity, as is the case in our implementation of
the algorithm, then the absolute free energy can be measured. The key
idea is that the normalization factor, defined in  Eq.~\eqref{eq:Q}, is
the ratio of the partition functions at the two inverse temperatures
$\beta$ to $\beta^\prime$. To see this, expand the definition of
$Z(\beta^\prime)$ and replace the resulting average at $\beta$ by its
population estimate,
\begin{eqnarray}
\label{nonlinear}
\frac{Z(\beta^\prime)}{Z(\beta)} 
&=& \frac{\sum_\gamma e^{-\beta^\prime E_\gamma}}{Z(\beta)}\nonumber\\
&=& \sum_\gamma e^{-(\beta^\prime-\beta) E_\gamma} ( \frac{e^{-\beta E_\gamma}}{Z(\beta)} )
\nonumber\\
&=& \langle e^{-(\beta^\prime-\beta) E_\gamma} \rangle_\beta
\nonumber\\
&\approx&  
\frac{1}{\tilde{R}_\beta} \sum_{j=1}^{\tilde{R}_\beta} e^{-(\beta^\prime-\beta) E_{\g}}=\wt(\beta,\beta^\prime) .
\end{eqnarray}
The summation over $\gamma$ in the first two lines in the above
expressions is over all possible spin configurations while the summation
in the last line is over the population.  Since $F = - T \ln  Z$, the
estimator of the free energy $\F$ at each simulated temperature is,
\begin{equation}
\label{eq:sumQ}
-\beta_k \F(\beta_k) 
= \sum^{N_T-1}_{\ell=k+1} \ln  \wt(\beta_{\ell},\beta_{\ell-1}) + \ln  \Omega ,
\end{equation}
where $\Omega$ is the number of microstates of the systems and $\{
\beta_\ell \}$ is the sequence of inverse temperatures in descending
order: $\beta_{N_T-1}=0$ and $\beta_0=1/T_0$ is the inverse of the
target temperature.  For Ising systems, $\Omega=2^N$ where $N$ is the
number of spins. Since the number of temperatures in population
annealing is typically in the hundreds, it is straightforward to
accurately measure free energy, energy and entropy as a continuous
function of temperature over the whole range of temperatures from
infinity to $T_0$.  It should be noted that the same method can also be
efficiently employed to measure free-energy differences in PT
\cite{wang:15b}.

\subsection{Weighted averages}
\label{sec:weightedaverage}

Many independent runs of PA for the same system may be combined to
reduce both systematic and statistical errors in the measurement of an
observable $\cal{A}$. Suppose we have carried out $M$ independent runs
and obtained estimates $\tilde{\cal A}_m$, $m=1,\ldots, M$.  Let
$\F_m(\bb)$ be the free energy estimated in run $m$ at the measurement
temperature $1/\beta$. If the different runs have different  population
sizes, let  $R_m$ be the nominal population size in run $m$.  Then, the
best estimator, $\overline{\cal A}$ for the thermal average of the
observable is,
\begin{equation}
\label{eq:weight}
\overline{\cal A} = \frac{\sum_{m=1}^M  \tilde{\cal A}_m R_m \exp[-\bb \F_m(\bb)]}{\sum_{m=1}^M R_m \exp[-\bb \F_m(\bb)]}.
\end{equation}
To justify this formula, consider an unnormalized variant of population
annealing in which the population is not kept under control but is
allowed to grow exponentially. In the resampling step in the
unnormalized version of PA, the expected number of copies of replica
$\g$ is simply the reweighting factor $\exp\left[
-(\beta^\prime-\beta)E_\g\right]$.  Unnormalized PA is equivalent to
standard PA except that it requires exponential computer resources and
yields better statistics. Without the normalization factor in the
resampling step, each replica evolves independently and combining
separate runs of the unnormalized algorithm requires no weighting factor
other than the obvious weighting by the population size, $R_m$.  Thermal
averages in unnormalized PA are obtained using simple averaging.  The
simple average in unnormalized PA becomes a weighted average in standard
PA because the populations in  different runs of standard PA have been
normalized differently.  Specifically, the product of the normalization
factors $\wt$ [Eq.~\eqref{eq:Q}] from the highest temperature to the
measurement temperature is the ratio of the population size in
unnormalized PA to the population size in standard PA.  But this product
is proportional to the exponential of the free energy [see
Eq.~\eqref{eq:sumQ}] justifying the use of $R_m \exp[-\beta \F_m(\bb)]$
as the weighting factor in standard PA. Note that observables such as
the spin and link overlap that involve more than one independent copy of
the system may also be estimated using weighted averages from multiple
independent runs as discussed in Sec.~\ref{sec:overlap}.

Weighted averaging for the dimensionless free energy is more complicated
because the free energy involves measurements at {\em all} temperatures,
however, as shown in Ref.~\cite{machta:10}, the final result is
relatively simple:
\begin{equation}
\label{eq:fweight}
-\bb \overline{F} = \ln  \left[ \frac{\sum_{m=1}^M   R_m \exp[-\bb \F_m]}{\sum_{m=1}^M R_m } \right] .
\end{equation}
This equation is obtained from Eq.~\eqref{eq:sumQ} and the fact that
$\wt(\beta_{\ell},\beta_{\ell-1})$ is an observable for which weighted
averaging applies, but at inverse temperature $\bb_\ell$.  Thus
\begin{equation}
\begin{aligned}
\label{eq:wxtsumQ}
&-\beta_k \overline{F}(\beta_k) = \\ 
& \sum^{N_T-1}_{\ell=k+1} \ln  \left[\frac{\sum_{m=1}^M  \wt_m(\beta_{\ell},\beta_{\ell-1}) R_m \exp[-\bb_{\ell} \F_m(\bb_{\ell})] }{\sum_{m=1}^M R_m \exp[-\bb_{\ell} \F_m(\bb_{\ell})]} \right]\\
 &\;\;\;\; + \ln  \Omega .
\end{aligned}
\end{equation}
This complicated equation for the weighted average of the dimensionless
free energy collapses to Eq.~\eqref{eq:fweight} after using the fact that
\begin{equation}
\wt_m(\beta_{\ell},\beta_{\ell-1})  \exp[-\bb_{\ell} \F_m(\bb_{\ell})] = \exp[-\bb_{\ell-1} \F_m(\bb_{\ell-1})],
\end{equation}
and also noting that the weighting factor at $\bb=0$ is simply $R_m$ and
setting $\beta_k=\bb$.

It is important to understand that combining multiple independent runs
with weighted averaging reduces both statistical errors {\em and}
systematic errors. By contrast, ordinary averaging reduces only
statistical errors. It is obvious that more measurements should reduce
statistical errors. Systematic errors are reduced because the weighted
average of multiple runs is identical to simulating a larger population
size and systematic errors diminish with population size. Indeed, all
ensemble averaged quantities are {\em exact} in the limit of an infinite
population size or, equivalently,  using weighted averaging in the limit
of an infinite number of runs with fixed population size.  If the
variance in $\beta \F(\beta)$ is much less than unity, there is little
difference between weighted averaging and simple averaging.  However, if
the variance of the free energy is large, the weighting factors, which
depend exponentially on the free energy, are broadly distributed, and
the two averages differ substantially.  As we shall see in the next
subsection, the variance of the free-energy estimator is a fundamental
quantity in understanding systematic errors in PA.

Note that there is no method available for combining independent runs of
a MCMC algorithm to decrease systematic errors. The most comparable
procedure to weighted averaging for MCMC algorithms is ``checkpointing.''
In checkpointing, the complete state of the system is saved at the end
of the simulation.  If results with smaller systematic errors are
required, the simulation can be restarted beginning with the final
state of the previous simulation so that averaging is initiated after a
longer initialization period. Compared to weighted averaging,
checkpointing requires substantially more storage because the full
configuration of the system must be stored, instead of just the
estimators for the observables and the free energy. In addition,
checkpointing must be done sequentially while weighted averaging can be
carried out using multiple parallel runs. It is a significant advantage
of PA that independent runs can be combined to improve systematic errors
(equilibration), which is {\em not} possible in PT.

\subsection{Macroscopic degrees of freedom}

Some problems in computational statistical mechanics require averaging
over a small discrete set of macroscopic degrees of freedom {\em in
addition} to a much larger number of microscopic degrees of freedom.  An
example of this situation is thermal boundary conditions for spin
models~\cite{wang:14}. For thermal boundary conditions in $d$ space
dimensions, the $2^d$ combinations of periodic and anti-periodic
boundary conditions in the $d$ directions are all included in the
thermal ensemble.  Each combination of spin configuration and boundary
condition appears in the ensemble with its Boltzmann weight.  Since
differing boundary conditions will have energies that differ by the
surface area of the system, energy differences are much greater than for
single spin flips.  Large energy differences between different boundary
conditions imply that Metropolis moves to change boundary conditions
will be strongly suppressed except at very high temperature.
Macroscopic degrees of freedom such as the boundary conditions in
thermal boundary conditions can be easily handled by PA if the starting
temperature in the simulation is $\bb=0$. At infinite temperature, each
macroscopic state appears with the same probability so the initial
population is set up with equal fractions in each macroscopic state. For
example, in three-dimensional ($d = 3$) Ising spin-glass simulations with
thermal boundary conditions, $1/8$ of the population is initialized in
each of the $2^3 = 8$ boundary conditions.  No Monte Carlo moves are
required to change boundary conditions during the remainder of the
simulation because the resampling step correctly takes care of adjusting
the fraction of each boundary condition in the population.  We have
successfully used this method to carry out large-scale simulations of
the Edwards-Anderson spin glass in thermal boundary
conditions~\cite{wang:14,wang:15a}. A more accurate but also more costly
method is to simulate each macroscopic state separately and then combine
them with weights given by the exponentials of their respective free
energies. Macroscopic degrees of freedom can also be efficiently
simulated using PT \cite{wang:15d}.

\section{Systematic and statistical errors}
\label{sec:error}

\subsection{Systematic errors and the variance of the free energy}
\label{sec:systematic}

Systematic errors in PA reflect the fact that for finite population size
$R$, the population is not an unbiased sample from the Gibbs
distribution. For PA, the algorithm `equilibrates' to the Gibbs
distribution as $R$ increases.  In this section we study the convergence
in $R$ to the equilibrium Gibbs distribution.  Consider the weighted
average of $M$ runs each with fixed population size $R$.  In what
follows a fixed value of $R$ is implicit in the notation.  We  argued in
Sec.~\ref{sec:weightedaverage} that the exact Gibbs ensemble average
$\langle {\cal A} \rangle$ of observable ${\cal A}$ is obtained by
weighted averaging in the limit of infinitely many runs,
\begin{equation}
\langle {\cal A} \rangle = \lim_{M \rightarrow \infty} \frac{\sum_{m=1}^M  \tilde{\cal A}_m  \exp[-\beta \F_m(\bb)]}{\sum_{m=1}^M  \exp[-\beta \F_m(\bb)]}.
\end{equation}
Replacing the sum over runs by an integral over classes of runs with
similar values of $\tilde{\cal A}$ and $\F$, we obtain,
\begin{equation}
\label{eq:A}
\langle {\cal A} \rangle= \frac{\int  \int   \ca \ p_{\AF}(\ca , \ff)  \exp[-\beta \ff] d \ca d\ff }{\int    p_F(\ff) \exp[-\beta \ff] d\ff},
\end{equation}
where $p_F(\cdot)$ is the probability density for free energy estimator
$\F$ and $p_{\AF}(\cdot , \cdot)$ is the joint probability density of
measuring observable $\tilde{\cal A}$ and free energy estimator $\F$.
The average of the estimator $\tilde{\cal A}$ in a single run of PA,
$\langle \tilde{\cal A} \rangle$ is
\begin{equation}
\label{eq:estA}
\langle \tilde{\cal A} \rangle =  \int   \ca \ p_{\cal A}( \ca) d \ca .
\end{equation}

Note that the difference between the integrals for $\langle {\cal A}
\rangle$ and $\langle \tilde{\cal A} \rangle$ is simply the weighting
factor $\exp[-\beta \F]$. The difference, $\Delta {\cal A} = \langle
\tilde{\cal A} \rangle - \langle {\cal A} \rangle$ is the systematic
error in measuring ${\cal A}$ in a single run of PA with population size
$R$.

Systematic errors for the free energy present a simpler situation.  The
cumulant generating function $\phi$ of $p_F$ is defined as,
\begin{equation}
\phi(z)=\ln  \left[ \int d \ff  \exp[z \ff] \,  p_F(\ff) \right].
\end{equation}
But $\phi(-\bb)$ is the integral expression for the weighted average of
the dimensionless free energy, see Eq.~\eqref{eq:fweight} with constant
$R_m$. Thus, the equilibrium free energy $F$ is related to the
distribution of the free energy estimator via,
\begin{equation}
F=    -\phi(-\bb)/\bb ,
\end{equation}
while the expected value of the free energy estimator from a single run,
$\langle \F \rangle$, is given by
\begin{equation}
\langle \F \rangle=  \frac{\partial}{\partial z}  \phi(z) \bigg |_{z=0} = \mu_F
\end{equation}
where $\mu_F$ is the mean of $p_F$. The systematic error in the free
energy is the difference between these expressions, $\Delta F= \langle
\F \rangle - F$.  Since  $\phi(z)$ is the cumulant generating function,
we see that
\begin{equation}
\label{eq:df}
\Delta F = \frac{1}{2} \bb \sigma_F^2 + \sum_{n=3}^\infty \frac{(-1)^n \bb^{n-1}}{n!} C_n,
\end{equation}
where $C_n$ is the n$^{th}$ cumulant of $p_F$ and $\sigma_F^2=C_2$ is
the variance of $p_F$.

For large population size ($R \gg 1$) an argument based on the central
limit theorem suggests that $p_F$ should become a Gaussian since $\F$ is
the sum of contributions from large number of nearly independent members
of the population. Thus, for large $R$ we expect the simpler expression,
\begin{equation} 
\label{eq:errF} 
\Delta F = \frac{\bb}{2} \sigma_F^2
\end{equation} 
to become exact.

Similarly, for large $R$ we expect the joint distribution $p_{\AF}$ in
Eq.~\eqref{eq:A} to be a bi-variate Gaussian defined by the means and
variances of $\tilde{\cal A}$ and $\F$, and their covariance, ${\rm
cov}(\tilde{\cal A},\F)$. Carrying out the Gaussian integrals for
$\langle {\cal A} \rangle$ in Eq.~\eqref{eq:A} we obtain for the
equilibrium value of the observable,
\begin{equation}
\langle {\cal A} \rangle =\mu_A - \bb \, {\rm cov}(\tilde{\cal A},\F).
\end{equation}
Thus, the systematic error $\Delta {\cal A} = \langle \tilde{\cal A}
\rangle - \langle {\cal A} \rangle$ in estimating the observable ${\cal
A}$ in a simulations with population size $R$ is given by
\begin{equation}
\label{eq:errAF}
\Delta {\cal A} =  \bb \, {\rm cov}(\tilde{\cal A},\F).
\end{equation}
We see that for large $R$, the systematic error in any observable is
proportional to the covariance of the observable with the free-energy
estimator.  This expression for the systematic error in ${\cal A}$ can
be rewritten in a form that emphasizes the central role of the variance
of the free energy,
\begin{equation}
\label{eq:errAFx}
\Delta {\cal A} =  
{\rm var}(\bb \F) \left[\frac{{\rm cov}(\tilde{\cal A},\bb \F)}{{\rm var}(\bb \F)}\right].
\end{equation} 
It is expected that the quantity in the square brackets will be nearly
independent of $R$ so that  systematic errors in ${\cal A}$ are
proportional to the variance of dimensionless free energy, just as is
the case for the free energy itself.

A central limit theorem argument suggests that ${\rm var}(\bb \F)$
decreases as $1/R$ so that the product $R \, {\rm var}(\bb \F)$ should
approach a constant. Define the {\em equilibration population size},
$\re$ as
\begin{equation}
\label{eq:re}
\re = \lim_{R \rightarrow \infty} R \ {\rm var}(\bb \F).
\end{equation}
The population is in equilibrium when $R$ is much larger than $\re$.
Define $\delta \tilde{\cal A} /\bb \delta \F$ as the limit of the
quantity in the square brackets in Eq.~\eqref{eq:errAFx}:
\begin{equation}
\frac{\delta \tilde{\cal A}}{\bb \delta \F} =  \lim_{R \rightarrow \infty} \frac{{\rm cov}(\tilde{\cal A},\bb \F)}{{\rm var}(\bb \F)}
\end{equation}
Given these definitions, the asymptotic theoretical prediction for
systematic errors is that,
\begin{equation}
\label{eq:error}
\Delta {\cal A} \sim  \frac{\re}{R } \frac{\delta \tilde{\cal A}} {\bb \delta \F} ,
\end{equation}
for any observable ${\cal A}$, except the free energy.  For the free
energy, the simpler expression holds:
\begin{equation}
\label{eq:ferror}
\Delta F =\frac{\re}{ 2 \bb R } .
\end{equation} 

Note that  $\delta \tilde{\cal A} / \delta \F$ can be interpreted as the
slope of the regression line through the joint distribution $p_{\AF}$.
To see this, let $\langle x\mid y \rangle$ be  the conditional average
of $x$ given $y$.  For a general bivariate normal distribution, the
conditional average is given by
\begin{equation}
\langle x\mid y \rangle  = \mu_x + \frac{{\rm cov}(x,y)}{\sigma_y^2}(y-\mu_y ),
\end{equation}
from which one sees that ${\rm cov}(\tilde{\cal A}, \F)/{\rm var}(\F)$
is the slope of the linear dependence of $\tilde{\cal A}$ on $\F$.  One
should not, however, consider Eq.~\eqref{eq:ferror} to be a special case
of Eq.~\eqref{eq:error} by setting $\delta \F / \delta \F=1$ since the
free-energy error equation has an extra factor of $1/2$.

For weighted averages, we expect similar results for systematic errors
but with $R$ replaced by $M R_0$, where $R_0$ is the size of the
individual runs and $M$ the number of runs in the weighted average.  The
substitutions $R \rightarrow M R_0$ in Eqs.~\eqref{eq:error} and
\eqref{eq:ferror} should become exact for weighted averages as $R_0/\re
\rightarrow \infty$ but for finite $R_0/\re$, where the joint
distribution is not close to a bivariate Gaussian, the dependence on $M$
may be more complicated.

\subsection{Statistical errors}
\label{sec:statisticalerrors}

The statistical error $\delta \tilde{\cal A}$ of an observable ${\cal
A}$ is the square root of the variance of the estimator
\begin{equation}
\delta \tilde{\cal A} \equiv [{\rm var}(\tilde{\cal A})]^{1/2}.
\end{equation}
Statistical errors scale inversely in the square root of the number of
independent observations.  In the absence of resampling, the number of
independent measurements in PA is the population size $R$.  However, the
resampling step makes identical copies of replicas and thus correlates
the population so that the effective number of independent measurements
is less than $R$.  On the other hand,  MCMC sweeps at each temperature
decorrelate the replicas.  Thus, if we consider only the correlating
effect of resampling we obtain an upper bound on the statistical errors.

Family trees can be constructed for each member of the initial
population.  Call all the descendants of replica $i$ in the initial
population a family and let $\fa_i$ be the fraction of the population
in family $i$. In a typical PA simulation starting at infinite
temperature and ending at a low temperature the great majority of
initial replicas have no descendants, $\fa_i=0$.   To obtain an upper
bound that ignores the decorrelating effect of the MCMC sweeps, assume
that observable ${\cal A}$ takes a single value $\tilde{\cal A}_i$ for
every member of family $i$.  If the MCMC algorithm applied at each
temperature step were completely ineffectual, this would be the case.
Given this assumption, the estimator $\tilde{\cal A}$ for the full
simulation is
\begin{equation}
\tilde{\cal A}  =  \sum_i  \fa_i \tilde{\cal A}_i .
\end{equation}
Next, make the additional approximation, which leads to an even weaker
upper bound, that the variance of the value of the observable in each
family is  ${\rm var}({\cal A})$, the full variance of the observable in
the thermal ensemble.  In particular, we are ignoring the possibility
that the observable is correlated with the family size.  For a given
distribution of family sizes, we obtain the variance of the estimator of
the observable ${\rm var}(\tilde{\cal A})$:
\begin{equation}
{\rm var}(\tilde{\cal A}) \leq {\rm var}({\cal A}) \sum_i  \fa_i^2 .
\end{equation}

Note that if every family contained one member and there were $R$
families then $\fa_i =1/R$ from which we would obtain the  result for
$R$ independent measurements, that $\delta \tilde{\cal A}= [{\rm
var}({\cal A})/R]^{1/2}$.  More generally, the statistical errors are
bounded by the second moment of the family size distribution.  Suppose
this moment scales as $1/R$ and define the {\em mean square family
size}, $\rt$:
\begin{equation}
\label{eq:rt}
\rt = \lim_{R \rightarrow \infty} R \ \sum_i  \fa_i^2.
\end{equation}
In terms of $\rt$, the bound on the statistical error in $\delta
\tilde{\cal A}$ is
\begin{equation}
\delta \tilde{\cal A} \leq \sqrt{{\rm var}({\cal A})\rt /R}.
\end{equation}
The quantity $R / \rt$ is an effective number of independent measurements.  

A second measure of the effective number of families is related to the
entropy $\Sf$ of the family size distribution
\begin{equation}
\Sf = -\sum_i \fa_i \ln  \fa_i .
\end{equation}
The exponential $e^{\Sf}$ is an effective number of families.   Suppose
$R/e^{\Sf}$ has a limit and define the {\em entropic family size},
$\rs$,
\begin{equation}
\label{eq:rs}
\rs = \lim_{R \rightarrow \infty} R /e^{\Sf}.
\end{equation}
The quantity $R/\rs$ is an alternative measure of the number of
independent measurements.  If every family is a singleton then
$\rs=\rt=1$.   If the family size distribution is exponential with mean
$\mu$ then it is straightforward to show that $\rt = 2 \mu$ and $\rs
\approx 1.53 \mu$.  As we shall see in Sec.~\ref{sec:rho} these two
measures are always close to one another.  All of the characteristic
sizes, $\re$, $\rs$ and $\rt$ are defined as limits as $R$ goes to
infinity but, in practice, we measure them at a fixed large $R$.

\subsection{Comparison of errors in population annealing and Markov-chain 
Monte Carlo algorithms}
\label{sec:compare}

In the previous two subsections we have seen that systematic and
statistical errors in PA both decrease with population size $R$;
systematic errors diminish as $1/R$, while statistical errors diminish
as $1/\sqrt{R}$. PA is a sequential Monte Carlo method while the great
majority of simulation methods in statistical physics are MCMC methods.
For MCMC methods observables are measured using time averages rather
than ensemble averages as is the case for PA and the equivalent quantity
to population size is the length of the run, ${\cal T}$. Errors are
reduced by increasing the running time and are estimated from the
autocorrelation functions of observables. Systematic errors in MCMC
diminish as $\exp(-{\cal T}/\tau_{\rm exp})$ where $\tau_{\rm exp}$ is
the ``exponential autocorrelation time,'' while statistical errors in an
observable ${\cal A}$ diminish as $\sqrt{2 \tau_{\rm int}^{\cal A}/{\cal
T}}$ where $\tau_{\rm int}^{\cal A}$ is the ``integrated autocorrelation
time'' for ${\cal A}$, see, for example, Ref.~\cite{janke:08} for a
discussion of integrated and exponential autocorrelation times.

In a loose sense we can equate the equilibration population size $\re$
with the exponential autocorrelation time $\tau_{\rm exp}$ and either of
the family size measures, $\rs$ or $\rt$, with integrated
autocorrelation times. Naively, it would appear that even if the
measures $\tau_{\rm exp}$ and $\re$ were comparable, a MCMC method would
have a considerable advantage over PA because MCMC algorithms converge
exponentially in the amount of computational work rather than inversely.
On further reflection, one can see that the exponential advantage of
MCMC is mostly illusory because of statistical errors, which decrease
only as the inverse square root of the amount of computational work for
both MCMC and PA. For both types of algorithms, the systematic errors
are dwarfed by the statistical errors for simulations of a single
system.

However, for disordered systems, it is usually necessary to carry out an
additional average over many realizations of the disorder.  Statistical
errors for disorder-averaged quantities decrease with the number of
disorder realizations, $n$ as  $1/\sqrt{n}$. When $n$ is large enough,
there could be a regime where statistical errors in disorder averages
are smaller than the systematic errors of PA. To investigate this issue
more quantitatively, consider an observable $\cal{A}$ and its disorder
average $\left[ \langle \cal{A} \rangle \right]_d$ where $\left[ \ldots
\right]_d$ indicates a disorder average. Using Eq.~\eqref{eq:error} we
have the following expression for the systematic error in the disorder
average, $\Delta \left[ \langle \cal{A} \rangle \right]_d$:
\begin{equation}
\Delta \left[ \langle \cal{A} \rangle \right]_d \approx  \left[     \frac{\re}{R } \frac{\delta \tilde{\cal A}} { \bb \delta \F}     \right]_d .
\end{equation}
Let $\delta \left[ \langle \cal{A} \rangle \right]_d$ be the statistical
error in $\left[ \langle \cal{A} \rangle \right]_d$ and suppose that the
main contribution to this statistical error comes from the  variance
with respect to disorder in  $\langle \cal{A} \rangle$ defined by
\begin{equation}
\Sigma_{\cal A}^2 = \left[ \langle {\cal A} \rangle^2 \right]_d- \left[ \langle \cal{A} \rangle \right]_d^2.
\end{equation}
Systematic errors are negligible relative to  statistical errors if
$\Delta \left[ \langle \cal{A} \rangle \right]_d  \ll  \Sigma_{\cal
A}/\sqrt{n}$ where $n$ is the number of disorder realizations in the
sample.  Thus, systematic errors are negligible if
\begin{equation}
\label{eq:diserror}
\frac{ \sqrt{n}}{  \Sigma_{\cal A}  }  \left[     \frac{\re}{R} \frac{\delta \tilde{\cal A}} { \bb\delta \F}     \right]_d  \ll 1,
\end{equation} 
and for the free energy we have the simpler expression,
\begin{equation}
\label{eq:fdiserror}
\frac{\sqrt{n}}{\Sigma_{\cal F}} \left[\frac{\re}{2 \bb R}\right]_d \ll 1,
\end{equation} 
In our $L=10$ simulations of the Edwards-Anderson model, discussed in
the following, $n=5000$ and $\Sigma_{F} = 23.9$ at $\bb=5$. Our
equilibration criterion requires that $\rs/R \leq 10^{-2}$ and, for most
instances, $\rs/R \ll 10^{-2}$. As we shall see in Sec.~\ref{sec:rho},
$\re$ is typically less than a factor of $2$ larger than $\rs$.  Thus,
the left-hand side of Eq.~\eqref{eq:fdiserror} for the disorder average
of the free energy is less than $10^{-2}$ and we are safely in the
regime where statistical errors  greatly exceed systematic errors.

\section{Model, Simulation Details, and Observables}
\label{sec:modelmethods}

\subsection{Edwards-Anderson model}

We test the performance of PA and compare it to PT in the context of the
three-dimensional (3D) Edwards-Anderson (EA) Ising spin glass model
\cite{edwards:75}, defined by the Hamiltonian
\begin{equation}
{\mathcal H} = - \sum_{\langle i,j \rangle} J_{ij} S_i S_j ,
\label{eq:ham}
\end{equation}
where $S_i \in \{\pm 1\}$ are Ising spins and the sum is over nearest
neighbors on a cubic lattice of linear size $L$ with periodic boundary
conditions. The random couplings $J_{ij}$ are chosen from a Gaussian
distribution with zero mean and unit variance. A set of couplings
${\cJ}=\{ J_{ij} \}$ defines a disorder realization or ``instance.''

Sampling low-temperature equilibrium states of the 3D EA model is
computationally very difficult. It is known that finding ground states
of the 3D EA models is an NP-hard computational problem
\cite{barahona:82} and it is believed that sampling low-temperature
equilibrium states is also exponentially hard in the sense that the
amount of computational work needed to achieve a fixed accuracy in
sampling grows exponentially in the system size $L$. For MCMC
algorithms, this intuition can be made more precise as a statement about
the $L$ dependence of autocorrelation times, while for PA it is a
statement about quantities such as $\re$, $\rt$ and $\rs$, introduced in
Sec.~\ref{sec:error}, which characterize population sizes required for
equilibration.

There are large sample-to-sample variations in the difficulty of
sampling equilibrium states of the 3D EA model. It is known that the
distribution of integrated autocorrelation times and other equilibration
measures for PT is approximately log-normal
\cite{katzgraber:06,alvarez:10a,yucesoy:13}. One of the important
question studied in Sec.~\ref{sec:results} is whether PA and PT both
find the same spin-glass instances to be either hard or easy.

There are  two  reasons why the 3D EA model is computationally difficult
that can be understood intuitively  in terms of the free-energy
landscape.  The first reason is that the free-energy landscape is rough
for typical instances with several relevant local minima separated by
high barriers. Both PT and PA are designed to partially overcome this
source of computational hardness though it certainly plays a
role~\cite{yucesoy:13}. The second reason is related to temperature
chaos~\cite{katzgraber:07,fernandez:13,wang:15a}, which is effectively a
change in dominance between minima in the free energy landscape as a
function of temperature. At high temperatures free-energy minima with
large entropies dominate while at lower temperatures free-energy minima
with low energies dominate and finding these {\em rare} low-energy
states is difficult for both PA and PT.

Extensive numerical evidence supports the idea that the 3D EA model
undergoes a second-order phase transition from a paramagnetic
high-temperature phase to a spin-glass phase at a temperature $T_c
\approx 0.96$ (see, for example, Ref.~\cite{katzgraber:06}).  Ordering
in the spin-glass phase is detected using the overlap distribution. The
overlap $q$ is defined as
\begin{equation}
\label{eq:q}
q= \frac{1}{N} \sum_i S_i^{(1)}S_i^{(2)},
\end{equation}
where $N=L^3$ is the number of spins, and the superscripts ``(1)'' and
``(2)'' refer to two statistically independent spin configurations
chosen from the Gibbs distribution with the same disorder $\cJ$. Let
$P_{\cJ}(q)$ be the overlap distribution for instance $\cJ$.  In the
paramagnetic phase and for large systems, $P_{\cJ}(q)$ is concentrated
near $q=0$, showing that independent spin configurations chosen from the
ensemble have little correlation.  The behavior of  $P_{\cJ}(q)$ for
large $L$ in the spin-glass phase is the subject of a longstanding
controversy but it is agreed that there are two peaks at $\pm q_{\rm
EA}$ with $q_{\rm EA}>0$ and $q_{\rm EA} \rightarrow 1$ as $T
\rightarrow 0$.  The controversy concerns whether or not $P_{\cJ}(q)$
simply consists of two delta functions at $\pm q_{\rm EA}$ as predicted
by the droplet picture \cite{mcmillan:84,fisher:86,bray:86} or whether
there is a forest of smaller $\delta$-functions between $-q_{\rm EA}$
and $+q_{\rm EA}$ as predicted by the replica symmetry breaking (RSB)
picture~\cite{parisi:79,mezard:87}. The droplet picture asserts that the
spin-glass phase consists of two thermodynamic pure states related by a
global spin flip while the RSB picture asserts that there exists a
countable infinity of thermodynamic pure states. For finite systems,
$P_{\cJ}(q)$ varies greatly from instance to instance with some disorder
realizations resembling the predictions of the droplet picture and
others the RSB picture.  The weight of $P_{\cJ}(q)$ near $q=0$ has been
used to distinguish the two competing theories of the low temperature
phase of the EA model.

\subsection{Simulation details}	

The large data sets used in this study were obtained in previous studies
of the low-temperature phase of spin glasses \cite{yucesoy:12,wang:14},
the dynamics of PT \cite{yucesoy:13}, and a comparison of PA and PT for
finding ground states \cite{wang:15}.  These data sets involve roughly
$n \approx 5000$ disorder realizations for each of five system sizes,
$L=4$, $6$, $8$, $10$, and $12$ (note that data for $L = 12$ using PT
have also been simulated, too, albeit at a higher temperatures).  The
same set of disorder realizations were simulated using both PA and PT to
allow for a detailed comparison between the two algorithms. In addition,
we carried out PA simulations for $n=1000$ instances with $L=14$.  The
parameters of the PA simulations are given in Table \ref{tab:parampa}. In
our implementation of PA, the annealing schedule has temperatures that
are evenly spaced in $\bb=1/T$ starting from infinite temperature. In
all PA simulations we used $N_S=10$ Metropolis sweeps per temperature.
The number of Metropolis sweeps per simulation, $\W$ is given by $\W=R
N_S N_T$ so that $\W$ is a rough measure of the computational work
expended per spin in the simulation.  For the $L=14$ runs we used
weighted averaging with $M=10$ independent runs per instance so that
here $\W=M R N_S N_T$.

The equilibration criterion that we use is that  $R \geq 100 \rho_s$, 
which is equivalent to $S_f \geq \ln (100)$.  It is worth mentioning 
here that $\rho_s$ converges rapidly as the population size grows (see
Fig. 10).  In our simulations, we first choose a population size for
which most instances are equilibrated. A larger population is used for
hard samples and the process is iterated until all samples meet the
equilibration criterion or until it becomes impractical to increase $R$.
In the latter case, we either use more temperatures or perform a 
weighted average.  For $M$ independent runs, it is straightforward to
show that the entropy of the family size distribution is given by
\begin{equation}
S_f=\sum\limits_{i=1}^{M} {S_{f,i}} p_i -\sum\limits_{i=1}^M p_i \ln(p_i) , 
\end{equation}
where $S_{f,i}$ is the entropy of the family size distribution for run 
$i$ and $p_i$ is the weight factor for run $i$, defined in Eq. 5.  Note
that if the $p_i$ and $S_{f,i}$ are both constants independent of $i$,
then from Eq. 31, $\rho_s$ is the same whether it is estimated from a
single long run with population $MR$ or $M$ shorter runs, each of length
$R$.

The population size for each system size is listed in the column labeled
$R$ in Table \ref{tab:parampa}. This population size satisfies the
equilibration criterion for most disorder realizations.  However, for
the hardest instances, runs were required with larger population sizes.
The number of hard instances, $n_{\rm hard}$ is listed in the last
column of the aforementioned table. The PA simulations were carried out
using {\texttt{OpenMP}} implemented on eight cores where each core works
on a different subset of the population. In addition to the simulations
described in Table \ref{tab:parampa}, we carried out a detailed study of
a single $L=8$ and a single $L=4$ disorder realization in which we
performed a large number of independent runs for various population
sizes to check predictions concerning systematic errors.

The parameters of the PT simulations are given in
Table \ref{tab:parampt}.  In the implementation of PT, the highest
temperature is $T=2$ and each PT sweep involves $N_S=1$ heat bath sweeps
per replica. Each simulation involves $2^{b+1}$ PT sweeps, $2^b$ for
equilibration and $2^b$ for data collection. The number of heat bath
sweeps per simulation and thus the computational work per spin is
$\W=2^{b+1} N_S N_T$.  In fact, for computing the overlap $q$, twice
this number of sweeps were used because two independent simulations are
needed to compute $q$ in PT. Additional details of the PT simulations
can be found in Ref.~\cite{yucesoy:13}.

\begin{table}
\caption{
Parameters of the main population annealing simulations~\cite{wang:14}.
$L$ is the system size, $R$ is the standard number of replicas, $T_0$ is
the lowest temperature simulated, $N_T$ the number of temperatures
(evenly spaced in $\bb$) in the annealing schedule, and $\W=R N_T N_S$
is the number of sweeps applied to a single disorder realization. $n$ is
the number of disorder realizations and $n_{\rm hard}$ is the number of
hard instances requiring more than $R$ replicas to meet the
equilibration requirement. For $L=14$ we used weighted averaging with
$M=10$ independent runs so $\W = M R N_T N_S$ for this case.
\label{tab:parampa}
}
\begin{tabular*}{\columnwidth}{@{\extracolsep{\fill}} l c  c c c c l l}
\hline
\hline
$L$  & $R$  & $T_0$ & $N_T$ & $\W$ & $n$ & $n_{\rm hard}$ \\
\hline
$4$  & $5\times10^4$ & $0.20$  & $101$ & $5\times10^7$  & $4941$ & $0$    \\
$6$  & $2\times10^5$ & $0.20$  & $101$ & $2\times10^8$  & $4959$ & $0$    \\
$8$  & $5\times10^5$ & $0.20$  & $201$ & $10^9$         & $5099$ & $5$    \\
$10$ & $       10^6$ & $0.20$  & $301$ & $3\times10^9$  & $4945$ & $286$  \\
$12$ & $       10^6$ & $0.333$ & $301$ & $3\times10^9$  & $5000$ & $533$  \\
$14$ & $3\times10^6$ & $0.333$ & $401$ & $1.2 \times 10^{10}$ & $1000$ & N/A \\
\hline
\hline
\end{tabular*}
\end{table} 

\begin{table}
\caption{
Parameters of the parallel tempering
simulations~\cite{yucesoy:12,yucesoy:13}. $L$ is the linear system size,
$2^b$ is the standard number of Monte Carlo sweeps.  $T_0$ is the lowest
temperature used, $N_T$ is the number of temperatures, and $\W=2^{b+1}
N_T N_S$ is the number of sweeps applied to a single disorder
realization.  $n$ is the number of disorder realizations.}
\label{tab:parampt}
\begin{tabular*}{\columnwidth}{@{\extracolsep{\fill}}  r c c c c c }
\hline
\hline
$L$ & $b$  & $T_0$ &  $N_{T}$& $\W$ & $n$ \\
\hline
$4$  & $18$ & $0.20$          & $16$ &  $8\times10^6$  & $4941$ \\
$6$  & $24$ & $0.20$          & $16$ &  $5\times10^8$  & $4959$ \\
$8$  & $27$ & $0.20$          & $16$ & $4\times 10^9$  & $5099$ \\
$10$ & $27$ & $0.20$          & $16$ & $4\times 10^9$  & $4945$ \\
\hline
\hline
\end{tabular*}
\end{table}

\subsection{Measured quantities}

We measured standard spin-glass observables and also quantities
intrinsic to the PA algorithm. We measured the internal energy
$\tilde{E}_{\cJ}$, free energy $\F_{\cJ}$, and spin overlap distribution
$\tilde{P}_{\cJ}(q)$ for all disorder realizations. From
$\tilde{P}_{\cJ}(q)$ we obtained its integral near the origin,
\begin{equation}
\tilde{I}_{\cJ} = \int_{-0.2}^{0.2} \tilde{P}_{\cJ}(q).
\end{equation}
From $\tilde{I}_{\cJ}$ for the $n$ instances we obtain the disorder
average $I=[\tilde{I}_{\cJ}]_d$. Unless required to prevent confusion,
we henceforth drop the subscript $\cJ$ indicating a particular instance.
Observables are obtained from population averages in contrast to the
situation for PT and other MCMC methods where observables are obtained
from time averages. Estimators of observables obtained from population
averages for a single instance are indicated by a tilde.

We estimated the family-based characteristic sizes, $\rt$ and $\rs$ for
each disorder realization.  For the $L=14$ and for the two individual
size $L=4$ and $L=8$ instances we also measured the equilibration
population size $\re$, which requires multiple runs.  These quantities
are defined as limits in $R$ but are estimated from the finite $R$
simulations.  Comparison data for PT were obtained in previous studies
\cite{yucesoy:12,yucesoy:13}. For the same set of disorder realizations,
we have values of $I_{\cJ}$ and the integrated autocorrelation time for
the spin overlap, $\tau_{{\cJ},{\rm int}}^{q}$.

\subsection{Spin overlap measurement}
\label{sec:overlap}

The spin overlap is an important quantity in spin-glass studies and its
integral near the origin, $I_{\cJ}$, has been extensively studied as a
way of distinguishing competing pictures of the low-temperature phase of
spin glasses. The measurement of the spin overlap distribution
$P_{\cJ}(q)$ would appear to be computationally twice as difficult as
other observables because it requires two independent spin
configurations. Indeed, in standard implementations of PT, two separate
simulations are run simultaneously and spin configurations from each are
combined to obtain values of $q$, so the work required to measure
$\tilde{P}(q)$ (and also the link overlap distribution
\cite{katzgraber:01}) is twice that for observables obtained from a
single spin configuration. In PA, however, it is possible to construct
$\tilde{P}(q)$ from a single run by taking advantage of the fact that
replicas from different families, {\em i.e.}, descended from different
initial replicas, are independent. We use the following method to
estimate $\tilde{P}(q)$ at a given temperature $\bb$.

First, a random permutation of the population, $( \pi_1, \pi_2, \ldots,
\pi_{\tilde{R}_\beta} )$ is constructed and used to make an initial
pairing of replicas in the population.  A random permutation is likely
to include pairs chosen from the same family.  If replica $k$ and
replica $\pi_k$ are in the same family, they are potentially correlated.
This ``incest'' problem is corrected sequentially by performing
transpositions as needed.  Suppose  $k$ is the least integer such that
replicas $k$ and $\pi_k$ are  in the same family.  Then the successive
replicas  $\pi_{k+1}, \pi_{k+2} \ldots $ are tested  until the first $j$
($j>k$) is found such that replica $\pi_j$ is in a different family than
replica $k$ and also replica $\pi_k$ is in a different family than
replica $j$.  The  permutation is now modified by transposing $\pi_j$
and $\pi_k$.  This process is continued until there are no more
incestuous pairs. Each pair then contributes one value to the histogram
for  $\tilde{P}(q)$.  Notice that in each step of the procedure the
number of incestuous pairs decreases by one.    So long as the maximum
family size is less than $\tilde{R}_\beta/2$, which is required anyway
for a well-equilibrated run,  this procedure will find an unbiased,
nonincestuous pairing.  Although the worst-case complexity of the
procedure is $O(R^2)$, in practice the complexity is
$O(R)$.

Weighted averaging may also be used to combine results for
$\overline{P}_{\cJ}(q)$ from many runs with $P_{\cJ}(q)$ playing the
role of the observable ${\cal A}$ in Eq.~\eqref{eq:weight}.  The
justification for weighted averaging based on unnormalized population
annealing holds, although the argument also requires the fact that each
family in unnormalized PA is independent and identically distributed.

\section{Results}
\label{sec:results}

In this section, we present results for both PA and PT.  This section
serves two purposes. The first purpose is to validate population
annealing and verify claims made in Sec.~\ref{sec:error} about its
statistical and systematic errors.  The second purpose is to compare the
efficiency of PA and PT.

\subsection{Spin overlap}

\begin{figure}[h]
\begin{center}
\includegraphics[scale=0.9]{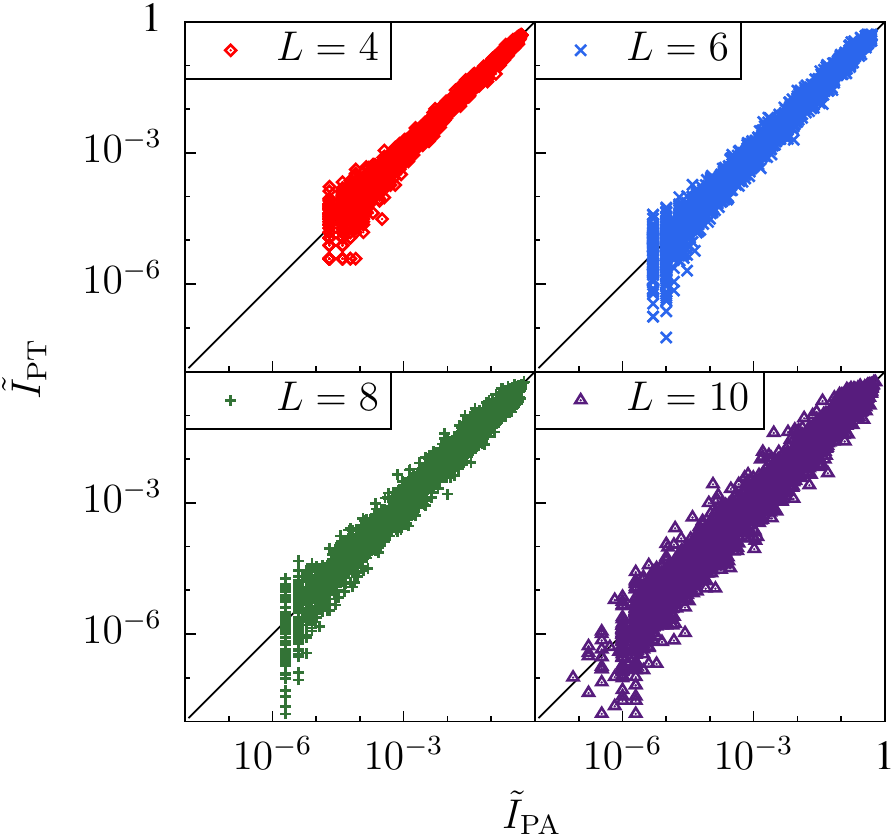}
\caption{(Color online)  
Log-log (base $10$) scatter plots of $\tilde{I}_{\cJ}$.  Each point
represents a disorder realization. The horizontal position of the point
is $\tilde{I}_{\cJ}$ measured in PA and the vertical position is the
value of $\tilde{I}_{\cJ}$ measured in PT, for sizes, $L=4$, $6$, $8$
and $10$ at $T=0.2$.}
\label{fig:ipapt}
\end{center}
\end{figure}

Figure \ref{fig:ipapt} shows a scatter plots for sizes $L=4$, $6$, $8$,
and $10$ of $\tilde{I}_{\cJ}$ for both algorithms, with the vertical
position of each point the value of $\tilde{I}_{\cJ}$ for PT and the
horizontal position the value for PA. Disorder realizations with
$\tilde{I}_{\cJ}=0$ for either algorithm are not shown. This figure
demonstrates reasonable agreement between the two algorithms for each
disorder realization. Note that PT is capable of measuring smaller
values of $\tilde{I}_{\cJ}$ than PA because the number of measurements
$2^b$ for PT is larger than the number of measurements $R$ for PA.

Next, we consider $I=[\tilde{I}_{\cJ}]_d$, the disorder average of the
integral of the spin overlap in the range from $-0.2< q <0.2$.  Table
\ref{tab:pzero} gives results for both PA and PT for $I$. The quoted
errors are obtained from the sample variance divided by the square root
of the sample size $\sqrt{n}$ so it is not surprising that the
difference between the PA and PT results is much less than the error
since both algorithms use the same set of disorder realizations.  It is
comforting that the results are so close. Because both algorithms are
quite different and use different criteria for equilibration it suggests
that systematic errors are minimal and cannot be detected in disorder
averages with a sample size of $5000$.

\begin{table}[h]
\caption{
Comparison of the disorder averaged overlap weight near the origin, $I$
between PA and PT at $T=0.2$ for the same set of disorder realizations.}
\label{tab:pzero}
\begin{tabular*}{\columnwidth}{@{\extracolsep{\fill}} c c c c c }
\hline
\hline
$L$ &4 &6 &8 &10 \\ \hline
PA &0.0186(10) &0.0194(10) &0.0205(10) &0.0200(10)\\ 
PT &0.0185(9) &0.0196(9) &0.0205(10) &0.0198(10) \\ 
\hline
\hline
\end{tabular*}
\end{table}

\subsection{Characteristic population sizes in PA and correlation times
in PT}
\label{sec:rho}

\begin{figure}[h]
\begin{center}
\includegraphics[scale=0.9]{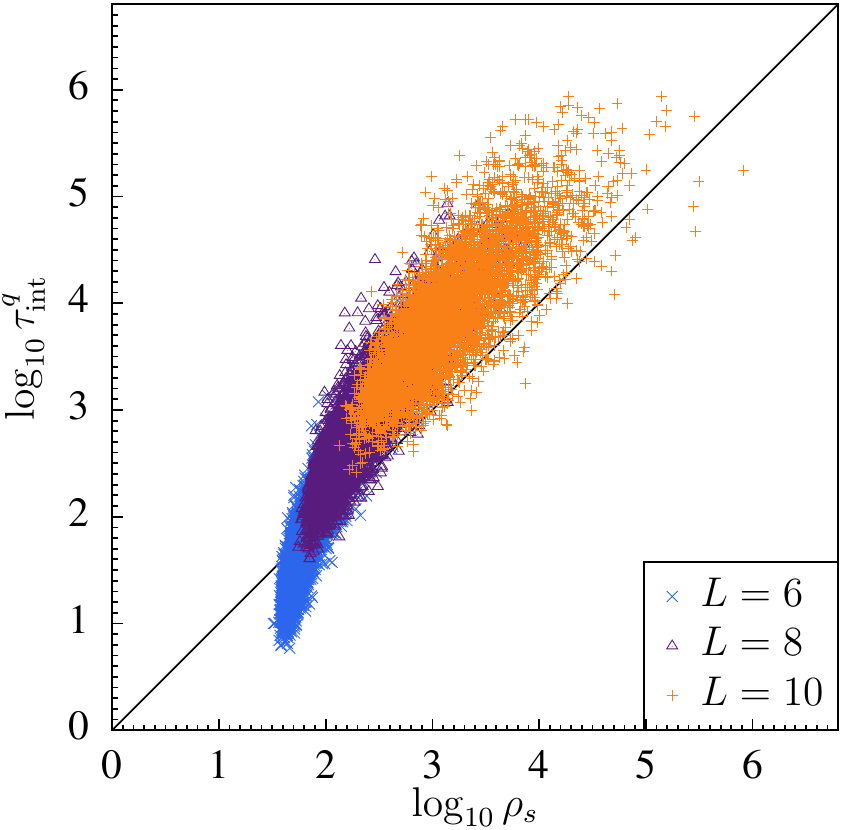}
\caption{(Color online)
Log-log scatter plot of $\rho_s$ (entropic family size for PA) vs
$\tau_{\rm int}^q$ (integrated autocorrelation time of the spin overlap
for PT). Each point represents a single disorder realization and there
are roughly $5000$ disorder realizations each for sizes  $L=6$, $8$, and
$10$ at $T=0.2$.}
\label{fig:tauvrho}
\end{center}
\end{figure}

Next we consider quantities that are intrinsic to each algorithm and
that are related to errors. Figure \ref{fig:tauvrho} is a logarithmic
scatter plot of $\rs$, the entropic family size measured in PA, and
$\tau^q_{\rm int}$, the integrated autocorrelation time for the spin
overlap measured in PT. Each point represents a disorder realization;
the horizontal position of the point is $\log_{10} \rs$ and the vertical
position is  $\log_{10} \tau^q_{\rm int}$. It is striking that the these
two quantities are strongly correlated. Both $\rs$ and $\tau^q_{\rm
int}$ are related to statistical errors in their respective algorithms
and large values correspond to hard instances that require lots of
computer resources to simulate accurately. It is clear that the
hardness of an instance for PA and for PT is strongly correlated.

\begin{figure}[h]
\begin{center}
\includegraphics[scale=.6]{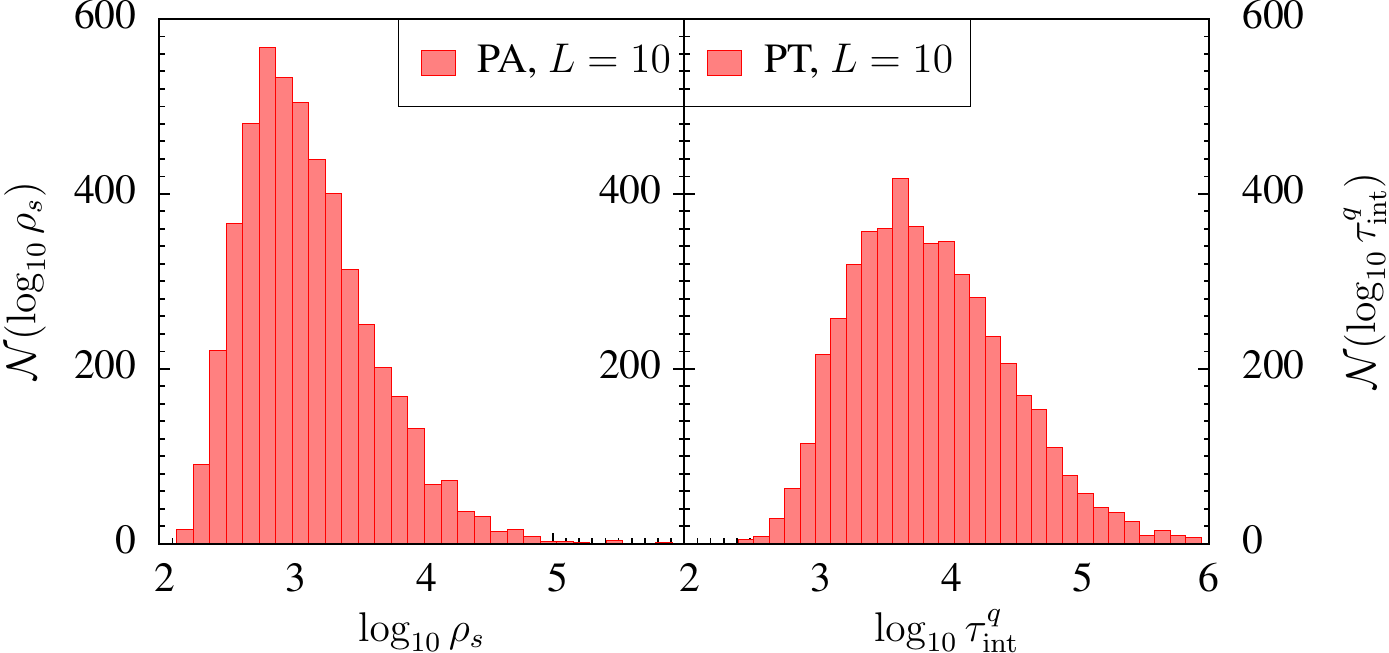}
\caption{(Color online) 
Histogram of $ \log_{10} \rs $  (left panel) and  $\log_{10} \tau^q_{\rm
int}$ (right panel) for all $4945$ disorder realizations, size $L=10$ at
$T=0.2$. }
\label{fig:rhotauhist}
\end{center}
\end{figure}

Figure \ref{fig:rhotauhist} shows histograms of $ \log_{10} \rs $ (left
panel) and $\log_{10} \tau^q_{\rm int}$ (right panel) for all $4945$
disorder realizations of size $L=10$ at $T=0.2$. Both distributions are
very broad and both are skewed toward hard disorder realizations although
the $\rs$ distribution is more sharply peaked than the $\tau^q_{\rm
int}$ distribution.

\begin{figure}[h]
\begin{center}
\includegraphics[scale=0.7]{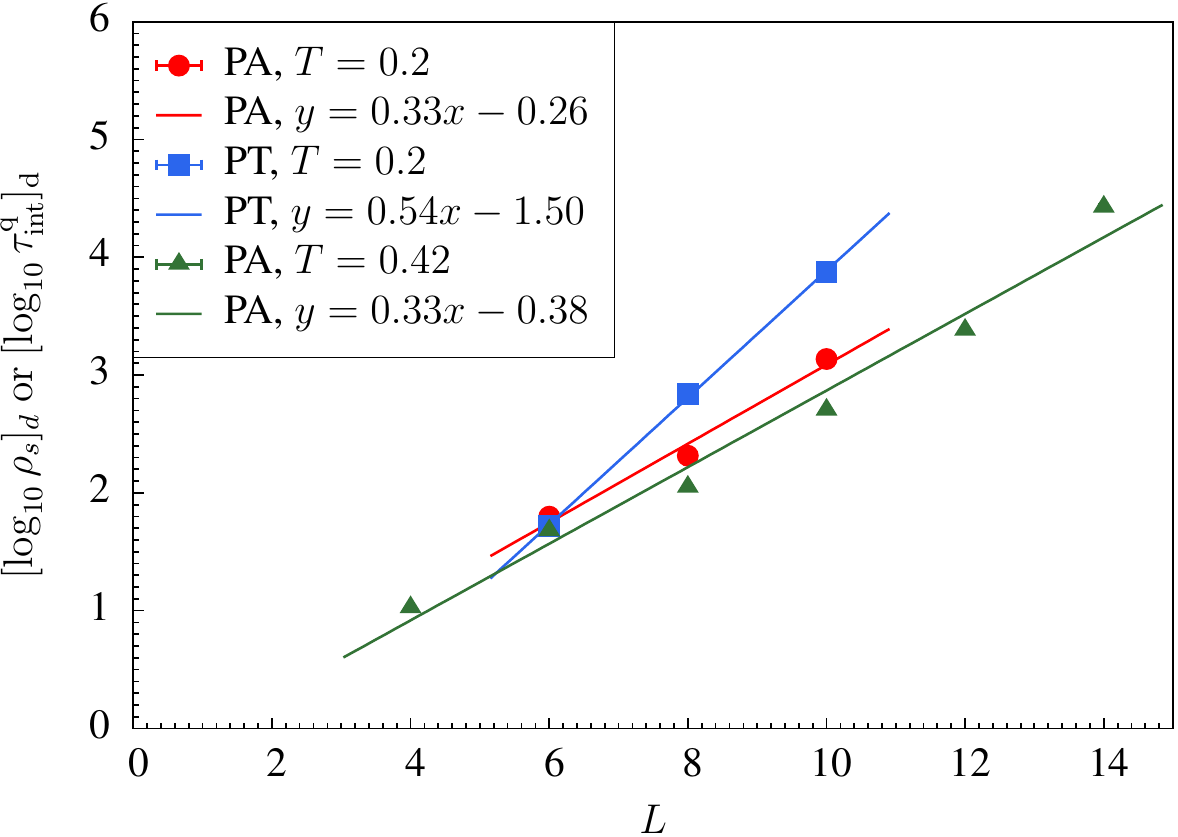}
\caption{(Color online) 
Disorder averages $[ \log_{10} \rs ]_d$ for PA and $[ \log_{10}
\tau^q_{\rm int} ]_d$ vs $ L$. Square symbols (blue) are for PT at
$T=0.2$, circular symbols (red)  for PA at $T=0.2$, and triangular
symbols (green) for PA at $T=0.42$. Straight lines are best linear fits
to the data. Note that the behavior of $\rt$ is qualitatively similar
to the behavior of $\rs$.}
\label{fig:rhotauvL}
\end{center}
\end{figure}

Figure \ref{fig:rhotauvL} is a log-linear plot of the disorder averages
$[\log_{10} \rs]_d$ and  $[\log_{10} \tau_{\rm int}]_d$ vs system size
$L$. Square symbols (blue) are for PT at $T=0.2$, circular symbols (red)
for PA at $T=0.2$, and triangular symbols (green) for PA at $T=0.42$. The
nearly linear behavior suggests that both algorithms suffer exponential
slowing with system size as expected. The fitted slope is greater for PT
than for PA, however, one should  be cautious in interpreting these fits
as indicating better scaling for PA relative to PT. There is some upward
curvature for PA in the data for both temperatures so the asymptotic
scaling slope may be significantly larger than the finite-$L$ slope. In
addition,  $\tau_{\rm int}$ and $\rs$ are not strictly comparable
quantities and, finally, neither algorithm has been carefully optimized.
Nonetheless, one can safely  conclude that PA is at least comparable in
efficiency to PT for the sizes studied -- system sizes that are of
current scientific interest across applications.

\begin{figure}[h]
\begin{center}
\includegraphics[scale=.8]{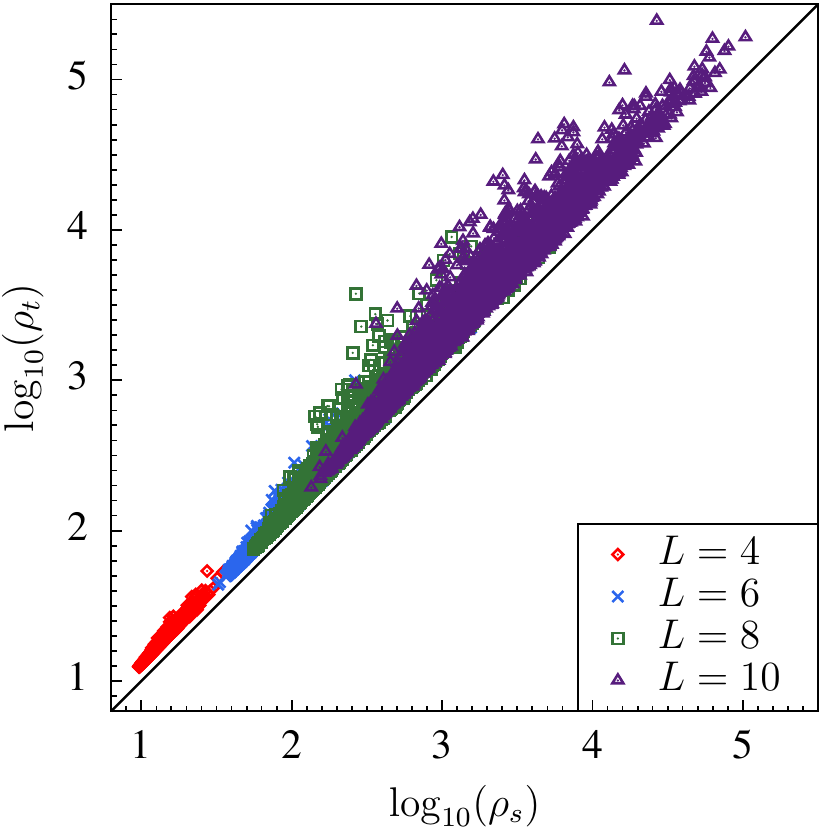}
\caption{(Color online) 
Scatter plot of $\rs$, entropic family size vs $\rt$ the mean squared
family size for sizes $L=4$, $6$, $8$, and $10$ at $T=0.2$.}
\label{fig:rhosvrhot}
\end{center}
\end{figure}

In Secs.~\ref{sec:systematic} and \ref{sec:statisticalerrors}, we
introduced three characteristic population sizes, $\rs$, $\rt$, and
$\re$. Both $\rs$ [see Eq.~\eqref{eq:rs}] and $\rt$ [see
Eq.~\eqref{eq:rt}] are obtained from the distribution of family sizes
and are related to statistical errors while $\re$ [see
Eq.~\eqref{eq:re}] is obtained from the variance of the free-energy
estimator and controls systematic errors.  What is the relation between
these three quantities for spin glasses? Figure \ref{fig:rhosvrhot} is a
scatter plot of $\rs$ vs $\rt$ for system sizes $L=4$, $6$, $8$, and
$10$.  Each point represents a single disorder realization. It is clear
that these two measures are strongly correlated with $\rs$ serving as a
lower bound for $\rt$.

\begin{figure}[h]
\begin{center}
\includegraphics[scale=0.8]{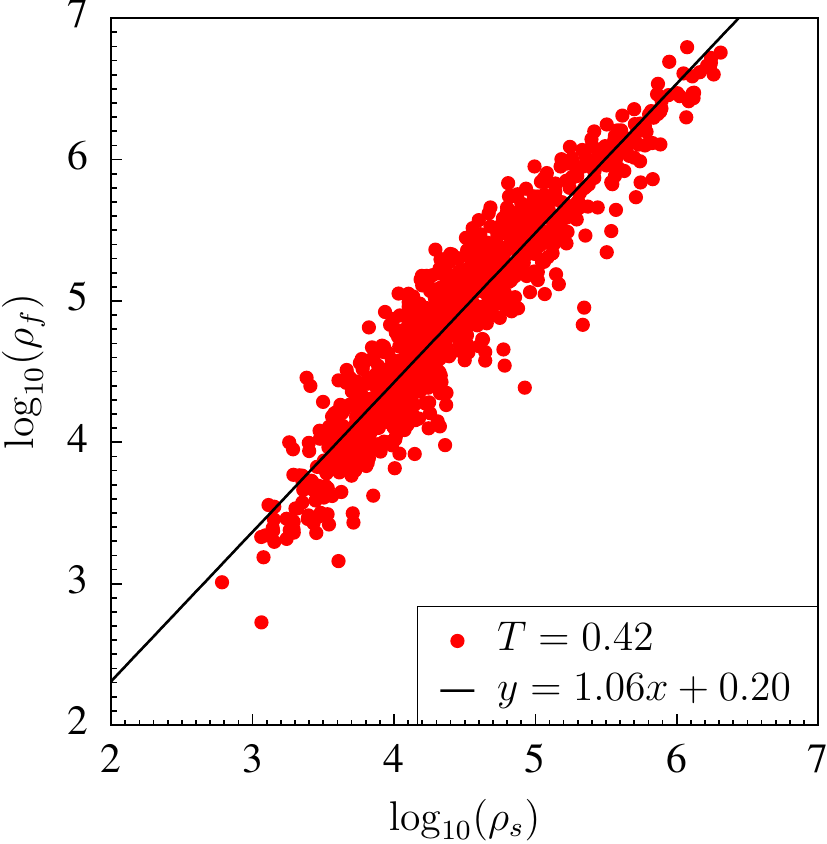}
\caption{(Color online)
Scatter plot of the entropic family size $\rs$ vs equilibration
population size $\re$ for $1000$ disorder realizations of size $L=14$ at
$T=0.42$. The straight line is a best fit through the data. }
\label{fig:rhosvrhof}
\end{center}
\end{figure}

Figure \ref{fig:rhosvrhof} is a scatter plot of $\rs$ vs $\re$ for the
$n=1000$ disorder realizations of size $L=14$ where each point
represents a disorder realization. The value of $\re$ is estimated for
each disorder realization from $10$ runs with $R= 3 \times 10^6$ and
$\re$ is estimated as $R$ times the sample variance of $\bb \F$ from the
$10$ runs. Since it is obtained from only $10$ runs, $\re$ has large
statistical errors. The straight line is a best fit through the data
points. It is clear that $\rs$ and $\re$ are strongly correlated
although $\re$ is on average a factor of $1.6$ larger than $\rs$.

The strong correlation between $\rs$ and $\re$ justifies using $R > 100
\rs$ as an equilibration criterion. In principle, equilibration
(systematic error) is controlled by $\re$ but measuring $\re$ requires
multiple runs whereas measuring $\rs$ requires only a single run.  Thus,
except for situations where weighted averaging is used, it is more
straightforward and reasonably well justified to require that the
population size is some factor larger than $\rs$. Because $\rt$ is just
as easy to measure as $\rs$, and $\rt>\rs$, and $\rt$ is more directly
related to statistical errors, it may be preferable to use $\rt$ rather
than $\rs$ as an equilibration criterion in future simulations.

\subsection{Convergence to equilibrium} 

Since statistical errors are much larger than systematic errors, in
order to investigate systematic errors, {\em i.e.},convergence to
equilibrium, it is necessary to carry out a very large number of
independent simulations of the same disorder realization. From these
many runs, systematic errors can be studied as a function of population
size $R$. In this section we examine in detail the convergence to
equilibrium for two disorder realizations. One of these disorder
realizations is the hardest $L=8$ sample as measured by $\rs$. This
disorder instance was also studied in detail in Refs.~\cite{wang:15} and
\cite{wang:15b}. We call this disorder realization ``instance \Y.'' The
second is an $L=4$ disorder realization that we call ``instance \Z."
Observables of the two instances are shown in Table
\ref{tab:sparameters}.

\begin{table}
\caption{
Equilibrium values of observables at $T=0.2$ for the two disorder
instances studied in detail, \Z\ and \Y, of sizes $L = 4$ and $8$,
respectively.
\label{tab:sparameters}
}
\begin{tabular*}{\columnwidth}{@{\extracolsep{\fill}} l c   c c c c  l l}
\hline
\hline
& $\re$ & $-\beta F$ & $-E$ & $\delta \tilde{E}/\bb \delta \F$ &$I$ &$\delta \tilde{I}/\bb \delta \F$\\
\hline
\Z  & $33$     & $584.138$ & $116.541$& $0.0542$ & $0.0929$&$0.0843$  \\
\Y  &  $9.0\times 10^3$     & $4457.53$ & $890.186$& $0.0355$ & $0.00104$& $0.00105$ \\ 
\hline
\hline
\end{tabular*}
\end{table} 

\begin{figure}[h]
\begin{center}
\includegraphics[scale=.65]{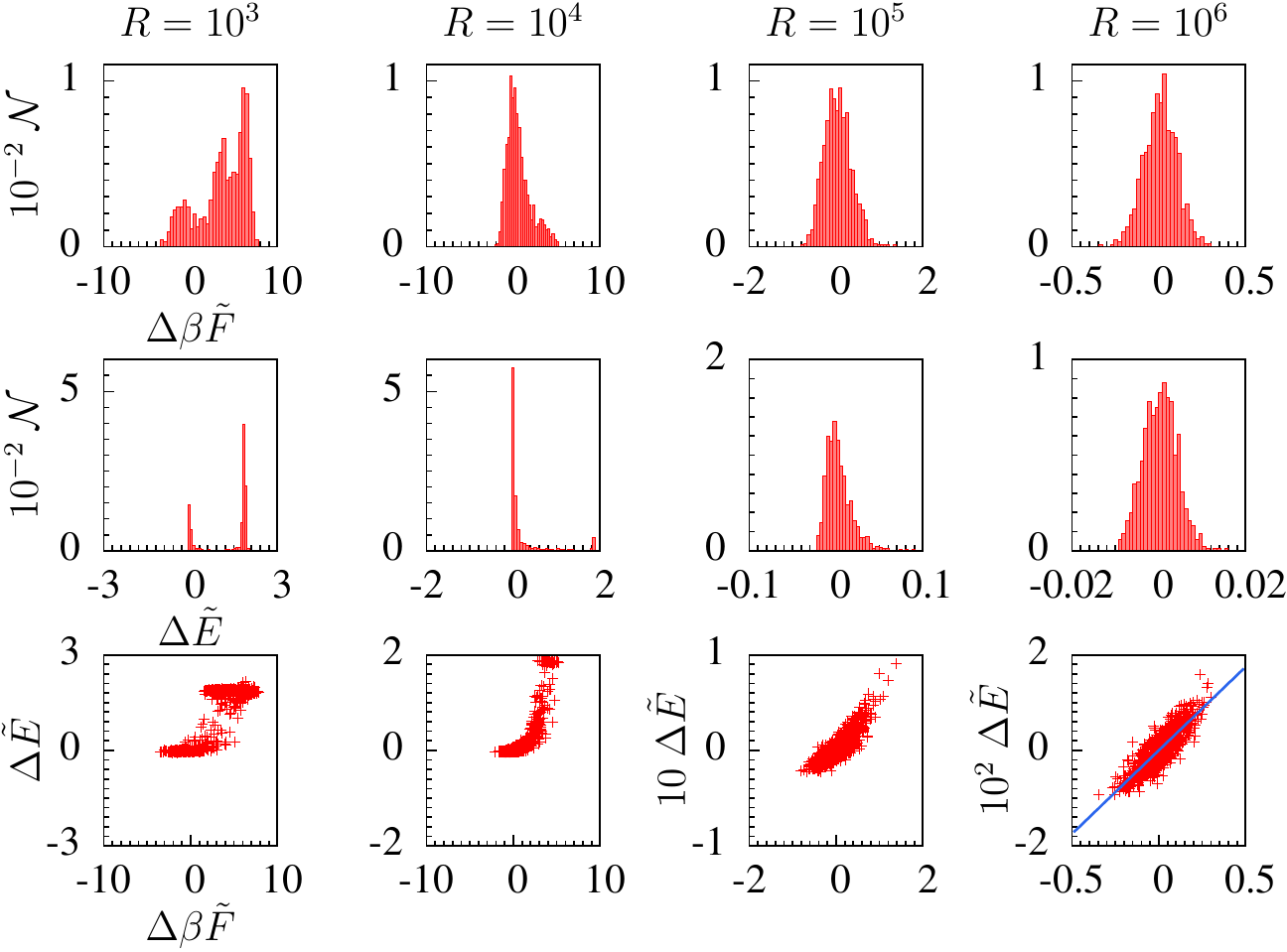}
\caption{(Color online) 
Histograms of $\Delta \bb \F$ (top row), histograms of $\Delta
\tilde{E}$ (middle row), and scatter plots representing the joint
distributions of $\Delta \tilde{E}$ and $\Delta \bb \F$ for instance \Y\
at $T=0.2$ (bottom row).  Each column represents a given population size
and, from left to right, $R=10^3$, $10^4$, $10^5$, and $10^6$,
respectively.  The slope of the regression line in the $\Delta
\tilde{E}$ vs $\Delta \bb \F$ scatter plot for $R=10^6$ (lower right
box) is the estimator of $\delta \tilde{E} /\bb \delta \F$. }
\label{fig:EFa}
\end{center}
\end{figure}

We first carefully examine, for instance \Y, the convergence to
equilibrium as a function of $R$ for the energy estimator $\tilde{E}$
and the dimensionless free-energy estimator $\bb \F$ at temperature
$T_0=0.2$. Figure \ref{fig:EFa} shows histograms of the deviation of the
dimensionless free-energy estimator from its equilibrium value, $\Delta
\bb\tilde{ F}$ (top row), the deviation of the  energy estimator from
its equilibrium value, $\Delta \tilde{E}$ (middle row), and a scatter
plot of their joint distribution (bottom row)  for population sizes,
$R=10^3$, $10^4$, $10^5$ and $10^6$ (from left to right, respectively).
For each population size we carried out ${\cal M} =1000$ independent
simulations of \Y. The `exact' equilibrium values, listed in Table
\ref{tab:sparameters}, are obtained from a weighted average of the
$1000$ runs at the largest population size, $R=10^6$. For the two
smaller populations the distributions are highly non-Gaussian but as $R$
increases the joint distribution approaches a bivariate Gaussian
distribution.  The joint distribution initially consists of two
well-separated peaks representing the fact for small $R$ most or all of
the population is frequently stuck in a metastable state with both a
higher free energy and higher energy. This bimodal distribution is a
feature of this particular disorder realization and explains, in part,
the computational hardness of this sample. Since $\re \approx 10^4$, the
$R=10^3$ populations are not equilibrated and the $R=10^4$ populations
are barely equilibrated. Finally, for $R=10^6$, the populations are
reasonably well equilibrated so that the $\tilde{E}$ and  $\bb \tilde{
F}$ distributions are close to Gaussian and the joint distribution is
close to a bivariate Gaussian.  The slope of the regression line through
the scatter plot representing the $R=10^6$ joint distribution is an
estimator of the quantity $ \delta \tilde{E} /\bb \delta \F$, which
controls the error in the energy estimator, see Eq.~\eqref{eq:error}.

We can assess more quantitatively  whether  $\bb \F$  and $\tilde{E}$
are described by a bivariate normal distribution.  From the ${\cal M}
=1000$ runs, we measured the skewness and kurtosis of both variables.
For instance, for \Y\ and $R=10^6$, the skewness and (excess) kurtosis
of the dimensionless free-energy runs is $0.047$ and $0.043$,
respectively.  Both values are statistically indistinguishable from
values that would be obtained from a sample of $1000$ normal random
variates.  The corresponding values of skewness and kurtosis for the
energy are $0.121$ and $0.152$, respectively.  Although larger, both
values are consistent with a sample of $1000$ normal random variates.
The joint distribution is, however, only marginally consistent with a
bivariate Gaussian, as measured by the Mardia \cite{mardia:70} combined skewness and
kurtosis test ($p=0.06$).  For instance, for \Y\ at $R=10^6$, $R/\re
\approx 10^2$. For \Z\ at population size $R=10^6$ we have $R/\re
\approx 3 \times 10^4$ and from ${\cal M} =5000$ runs the joint
distribution cannot be distinguished from a bivariate Gaussian by the
Mardia combined test ($p=0.4$).

\begin{figure}[h]
\begin{center}
\includegraphics[scale=.7]{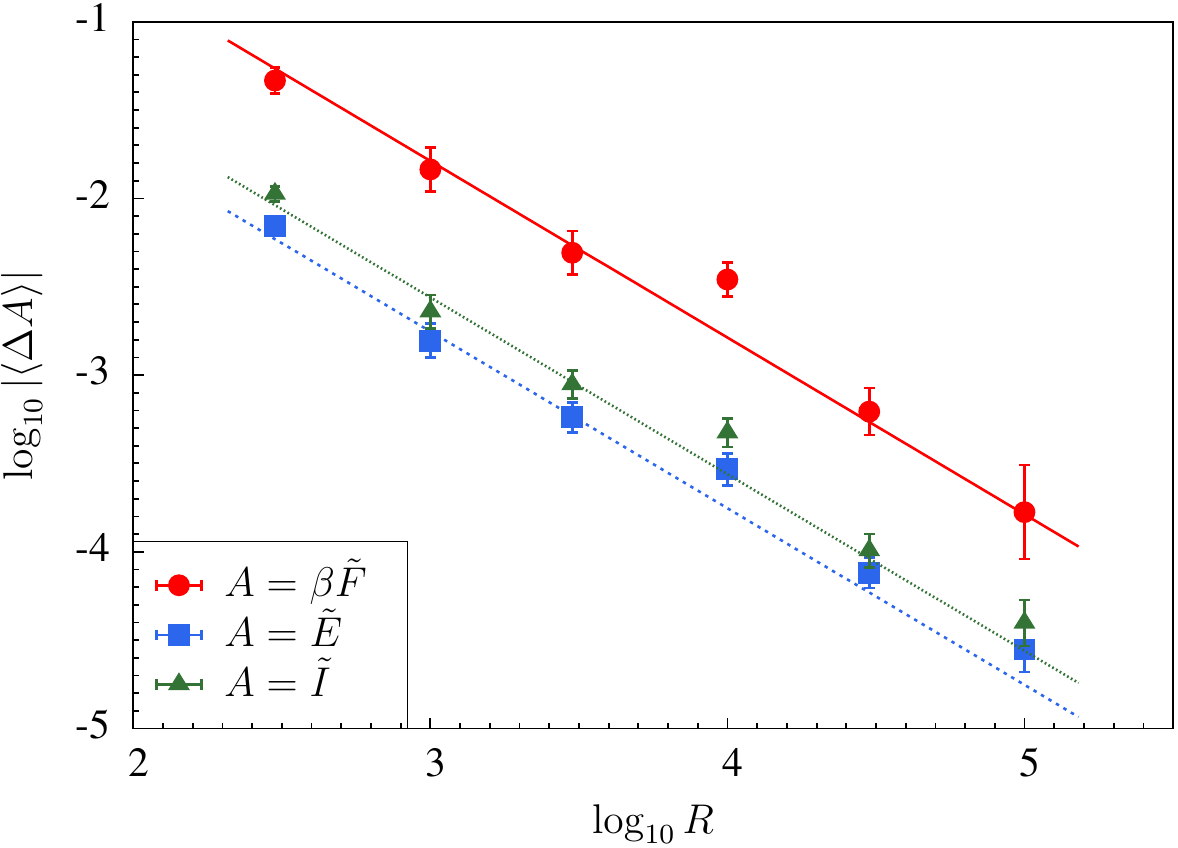}
\caption{(Color online)  
Log-log plot showing the deviations from equilibrium (systematic errors)
in the dimensionless free energy, $\langle \Delta \bb \F \rangle$ (red
circles),  energy, $\langle  \Delta \tilde{E} \rangle$ (blue squares)
and overlap near the origin $\langle \Delta \tilde{I} \rangle$ (green
triangles) as a function of population size $R$ for instance \Z \ at
$T=0.2$.  The straight lines are theoretical curves based on
Eqs.~\eqref{eq:error} and \eqref{eq:ferror}.}
\label{fig:avsr4}
\end{center}
\end{figure}

Next, we study the convergence of the mean values of observables to
their equilibrium values as a function of $R$. For each observable
${\cal A}$ we obtain the mean value for a single run $\langle
\tilde{\cal A} \rangle$ from a simple average over all ${\cal M}$ runs
for each population size and we obtain the equilibrium value from a
weighted average over all runs at the largest size.  Figure
\ref{fig:avsr4} shows $\langle\Delta \bb  \F \rangle$,  $\langle \Delta
\tilde{E} \rangle$, and $\langle \Delta  \tilde{I} \rangle$, the
deviation of the estimators of the dimensionless free energy, energy and
overlap near the origin from their respective equilibrium values, as a
function of population size $R$ for instance \Z. The straight lines are
theoretical curves from Eqs.~\eqref{eq:error} and \eqref{eq:ferror}
using the values of $\re$ and $\delta \tilde{\cal A}/\bb \delta \F$
estimated at $R=10^6$ and given in Tab.~\ref{tab:sparameters}. We see
that there is reasonable quantitative agreement with the predicted $1/R$
dependence of the systematic errors. The $R=10^6$ data point is not
shown because statistical errors in measuring the exact values $\langle
{\cal A} \rangle$ are comparable here to systematic errors in $\langle
\tilde{\cal A} \rangle$. Probing the $1/R$ regime of systematic errors
proved quite difficult because of the much larger statistical errors.
For example, to sufficiently reduce statistical errors for instance \Z\
we used ${\cal M}=32\,000$ independent runs to obtain the $R=10^5$
averages $\langle \tilde{\cal A} \rangle$ in Fig.~\ref{fig:avsr4} and
$M=5000$ independent runs at $R_0=10^6$ to obtain  ``exact'' equilibrium
values $\langle {\cal A} \rangle$ from weighted averaging.

\begin{figure}[h]
\begin{center}
\includegraphics[scale=.7]{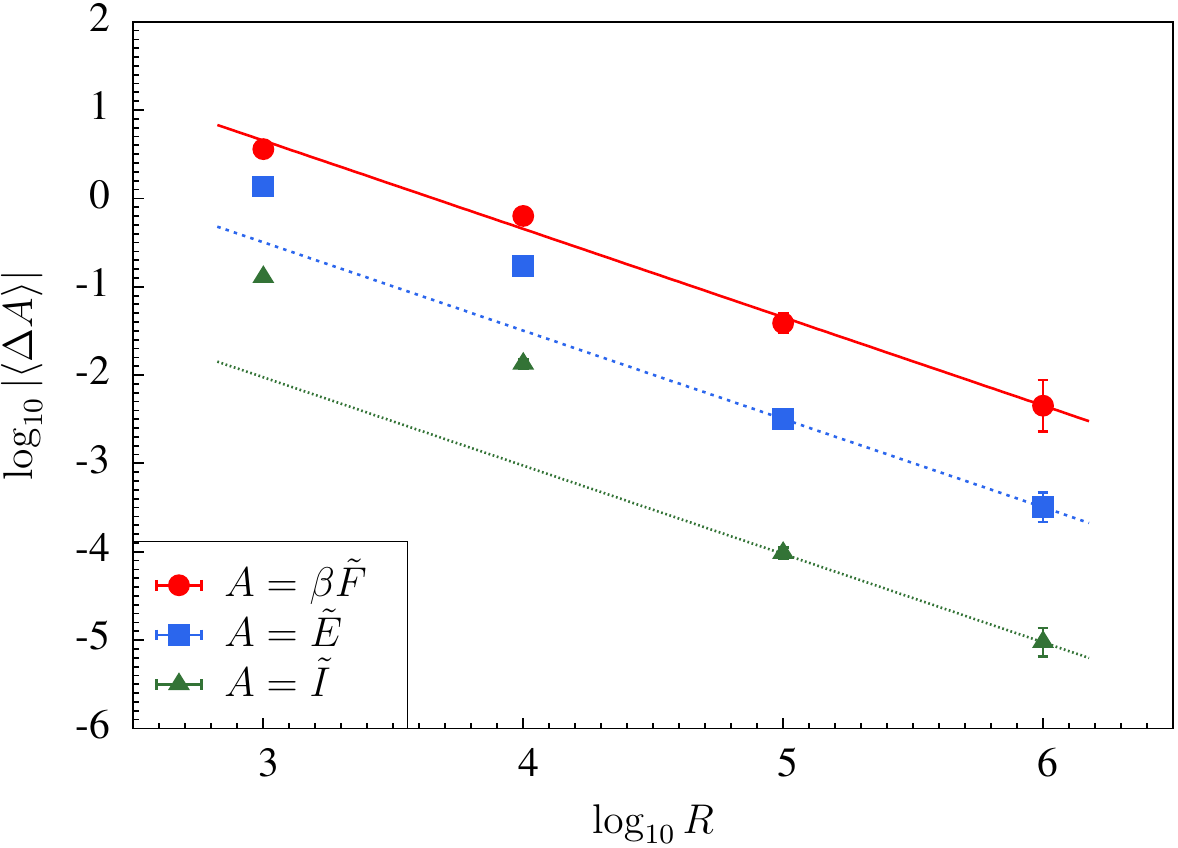}
\caption{(Color online) 
Log-log plot showing the deviations from equilibrium (systematic errors)
in the dimensionless free energy $\langle \Delta \bb \F \rangle$ (red
circles),  energy, $\langle  \Delta \tilde{E} \rangle$ (blue squares),
and overlap near the origin $\langle \Delta \tilde{I} \rangle$ (green
triangles) as a function of population size $R$ for instance \Y \ at
$T=0.2$. The straight lines are theoretical curves based on
Eqs.~\eqref{eq:error} and \eqref{eq:ferror}. }
\label{fig:avsr8}
\end{center}
\end{figure}

Figure \ref{fig:avsr8} shows similar results for the size $8$ instance
\Y. Since the joint distributions are far from bivariate Gaussians for
the smaller values of $R$ for instance \Y, the theoretical  predictions
for $\langle \Delta  \tilde{E} \rangle$, and $\langle \Delta  \tilde{I}
\rangle$  are poor for the smaller population sizes.  The  points for
$R=10^6$ in Fig.~\ref{fig:avsr8} are in essentially perfect agreement
with the theoretical predictions of Eq.~\eqref{eq:error},  however,
since $\re$ and $ \delta \tilde{\cal A} / \bb \delta \F$ are all
measured at $R=10^6$  this agreement is really just a check that the
joint distribution is close to the assumed bi-variate Gaussian form.

\begin{figure}[h]
\begin{center}
\includegraphics[scale=.7]{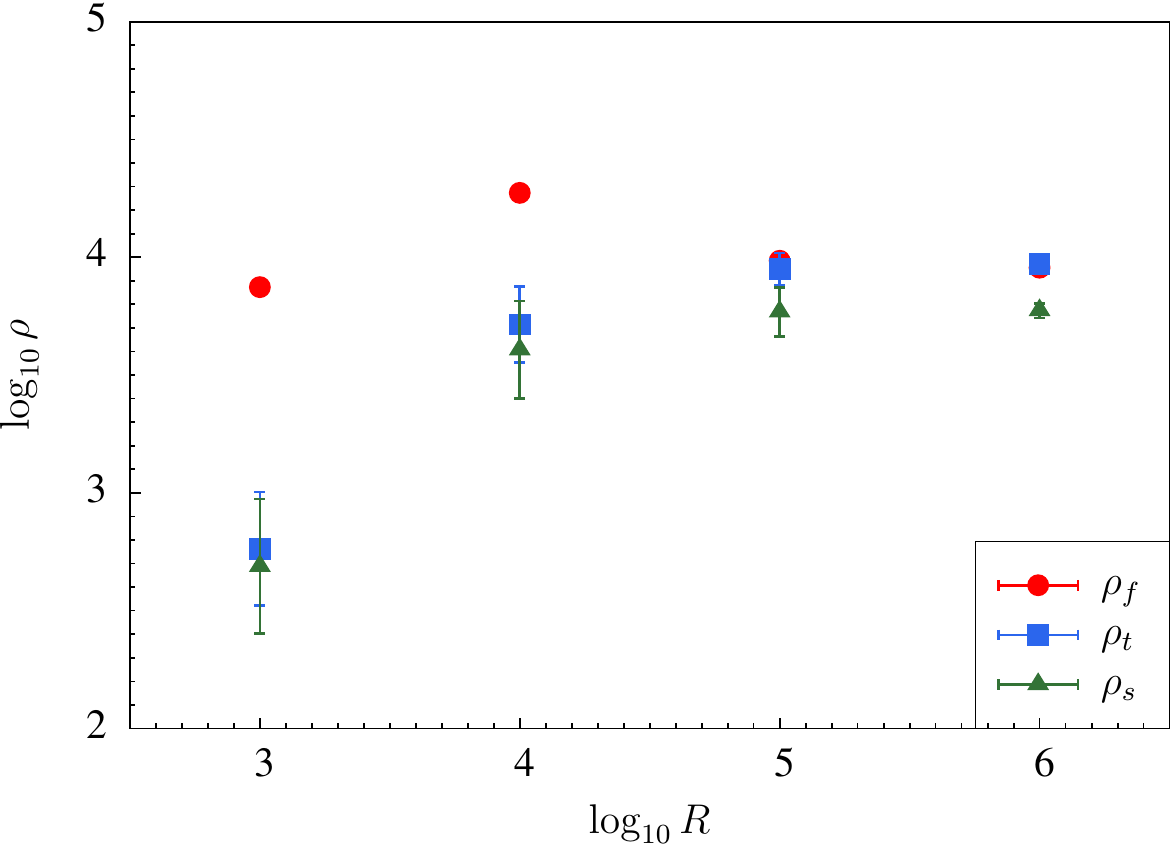}
\caption{(Color online) 
Log-log plot showing estimators of the equilibration sizes $\re$ (red
circles), $\rt$ (blue squares), and $\rs$ (green triangles) as a function
of population size $R$ at  $T=0.2$ for instance \Y.}
\label{fig:rhos}
\end{center}
\end{figure}

Next, we examine the convergence of the various characteristic
population sizes to their asymptotic values. Figure \ref{fig:rhos} shows
the finite size estimators of $\re$, $\rs$, and $\rt$ versus the population
size $R$ at which they are measured for instance \Y.  For this instance
all of these quantities have values near $10^4$ and their values are
near their asymptotic values for the two largest population sizes for
which $R \geq 10 \rho$.  The rapid convergence of $\re$ supports the
hypothesis that equilibrium is approached as $1/R$.  We do not show a
similar graph for instance \Z\ because all three $\rho$ measures are
already saturated to their asymptotic values within statistical errors
even for smallest population sizes studied.

Finally, we can also gain some insights into weighted averaging from the
detailed study of a single instance. The question we address is whether
a single run is significantly better than a weighted average with the
same total population size. We computed the systematic error in the
weighted average of the dimensionless free energy $\bb \F$ of instance
\Y\ for $R_0=10^3$ and $M=10$ and compared it to the systematic error
for a single run with $R=MR_0=10^4$.  We used ${\cal M}=1000$
independent runs with $R=10^3$ to compute  the mean $\langle \bb
\overline{F} \rangle$ and standard error of the weighted average. To
compute the mean, we take $M=10$ random values from the set of ${\cal
M}=1000$ runs, compute the weighted average, and then take the mean of
that weighted average over many such experiments. We used the blocking
method to compute the standard error of the mean. We used ${\cal
M}=1000$ runs with $R=10^4$ to obtain $\langle \bb \F \rangle$ and its
error.  We found that $\langle \Delta \bb \overline{F} \rangle= 0.75 \pm
0.12$, while $\langle \Delta \bb \F \rangle= 0.63 \pm 0.04$.  Thus, the
weighted average has roughly the same systematic errors as the single
long run.  Note that in this example,  $R_0 < \re$.  We expect the
differences between weighted averaging and a single long run to vanish
as $R_0/\re \rightarrow \infty$.  Unfortunately, even with $1000$
independent runs, we did not achieve sufficient statistical power to
distinguish the weighted average clearly from a single long run although
we expect the former to have somewhat larger systematic errors.  These
considerations lead to the following conjecture: Suppose one has
available a fixed total amount of work defined by a total population
size,  $R_0 M$ such that $R_0 \gtrsim \re$, then the weighted average
obtained from $M$ runs each with population size $R_0$ is statistically
indistinguishable from a single long run with population $R=R_0 M$.
However, the discussion in Sec.~\ref{sec:compare} comparing PA and PT
for disorder averaging is relevant here as well.  If a sufficiently
large disorder sample is simulated, the differences in systematic errors
between a weighted average and a single long run could become relevant.
While additional work to understand the systematic errors in weighted
averaging is needed, it seems clear that weighted averaging is a useful
tool for studying hard problems requiring very large total populations.

\section{Discussion}
\label{sec:discussion}

We have shown that population annealing is an effective and efficient
algorithm for simulating spin glasses. It is comparably efficient to
parallel tempering, the standard in the field, and it has several
advantages. 

The first advantage is that it is naturally a massively parallel
algorithm. The convergence to equilibrium occurs as the population size
grows and each replica in the population can be simulated independently.
Since realistic spin-glass simulations using population annealing
require population sizes of the order of $10^6$ or more, there is a much
greater scope for parallelism than in parallel tempering where in
spin-glass simulations typically less than $100$ replicas are simulated
in parallel. To put this difference in perspective, recall that parallel
tempering is a Markov-chain Monte Carlo algorithm while population
annealing is a sequential Monte Carlo algorithm. From a computational
complexity perspective, when going from a Markov-chain Monte Carlo
algorithm to a sequential Monte Carlo algorithm, time is exchanged for
hardware so that long running times can be exchanged for massive
parallelism. The downside of exchanging time for hardware is  that
population annealing has much larger memory requirements than parallel
tempering.

A second advantage of population annealing is access to weighted
averaging, which allows multiple independent runs of PA to be combined
to improve both statistical {\em and} systematic errors. Weighted
averaging opens the door to distributed computing. It is potentially
possible to quickly simulate very difficult to equilibrate instances of
spin glasses or other hard statistical-mechanical models by distributing
the work over a large and inhomogeneous set of computational resources.
The only information that needs to be collected and analyzed centrally
from each run is the estimators of observables together with the
estimator of the free energy.

Apart from its large memory usage, the main disadvantage of population
annealing (and sequential Monte Carlo methods in general) is that it
converges to equilibrium inversely in population size whereas parallel
tempering (and Markov-chain Monte Carlo methods in general) converges
exponentially. In most situations, this difference is moot because
statistical errors are much larger than systematic errors. However, for
very high precision disorder averages, it is possible that the
exponential convergence of parallel tempering would be an advantage over
the power law convergence of population annealing.

Thus far, the implementations of population annealing for large-scale
simulations have used a simple annealing schedule.  The temperature set
is uniform in the inverse temperature and there are a constant number of
Metropolis sweeps at each temperature. It is plausible that a more
complicated annealing schedule might be more efficient. It is perhaps
possible that the annealing schedule can be adaptively adjusted to the
particular problem instance in analogy to related proposals for parallel
tempering simulations \cite{katzgraber:06a,bittner:08}. It might also
improve efficiency to change the population size with temperature. In
addition, our implementation uses the Metropolis algorithm at every
temperature, however, at low temperatures kinetic Monte Carlo might be
preferable and, at intermediate temperatures cluster moves, might be
useful \cite{zhu:15}.

Population annealing is a general method suitable for simulating
equilibrium states of systems with rough free-energy landscapes.  It can
be applied to any system for which there is a parameter, such as
temperature, that takes the equilibrium distribution from an easy to
simulate region, {\em e.g.}, at high temperature, to a hard to simulate
region, {\em e.g.}, at low temperature. In addition to spin systems,
population annealing may prove useful in simulating the equilibrium
states of dense fluids or complex biomolecules.

\acknowledgments

The authors acknowledge contributions from Burcu Yucesoy in providing
parallel tempering data and from Jingran Li who participated in early
studies of population annealing. J.~M.~and W.~W.~acknowledge support
from National Science Foundation (Grant No.~DMR-1208046).
H.~G.~K.~acknowledges support from the National Science Foundation
(Grant No.~DMR-1151387) and would like to thank Dan Melconian and his
beloved friend Zaya for inspiration. H.~G.~K.'s research is in part
based upon work supported in part by the Office of the Director of
National Intelligence (ODNI), Intelligence Advanced Research Projects
Activity (IARPA), via MIT Lincoln Laboratory Air Force Contract
No.~FA8721-05-C-0002.  The views and conclusions contained herein are
those of the authors and should not be interpreted as necessarily
representing the official policies or endorsements, either expressed or
implied, of ODNI, IARPA, or the U.S.~Government. The U.S.~Government is
authorized to reproduce and distribute reprints for Governmental purpose
notwithstanding any copyright annotation thereon.  We thank the Texas
Advanced Computing Center (TACC) at The University of Texas at Austin
for providing HPC resources (Stampede cluster), and Texas A\&M
University for access to their Ada, Curie, Eos and Lonestar clusters.

\bibliographystyle{apsrevtitle}
\bibliography{refs}

\end{document}